\PassOptionsToPackage{top=3cm,left=3cm,right=3cm,bottom=3cm}{geometry,appendix}
\documentclass[fleqn,11pt]{wlscirep}

\usepackage{import}
\usepackage{main}
\usepackage{mathabx}

\begin{document}
\nolinenumbers
\doublespacing

\title{\centering\LARGE\singlespacing{A comparison of short-term probabilistic forecasts for the incidence of COVID-19 using mechanistic and statistical time series models}}

\author[1,2]{Nicolas Banholzer}
\author[3]{Thomas Mellan}
\author[3]{H Juliette T Unwin}
\author[4,5]{Stefan Feuerriegel}
\author[6]{Swapnil Mishra}
\author[3,7,*]{Samir Bhatt}

\affil[1]{ETH Zurich, Department of Management, Technology, and Economics, Zurich, Switzerland}
\affil[2]{University of Bern, Institute of Social and Preventive Medicine, Bern, Switzerland}
\affil[3]{Imperial College London, School of Public Health, London, United Kingdom}
\affil[4]{LMU Munich, LMU Munich School of Management, Munich, Germany}
\affil[5]{Munich Center for Machine Learning (MCML), Munich, Germany}
\affil[6]{National University of Singapore and National University Health System, School of Public Health and 
Institute of Data Science, Singapore}
\affil[7]{University of Copenhagen, Department of Public Health, Copenhagen, Denmark}
\affil[*]{Corresponding author: samir.bhatt@sund.ku.dk}

\begin{abstract}\normalfont
Short-term forecasts of infectious disease spread are a critical component in risk evaluation and public health decision making. While different models for short-term forecasting have been developed, open questions about their relative performance remain. Here, we compare short-term probabilistic forecasts of popular mechanistic models based on the renewal equation with forecasts of statistical time series models. Our empirical comparison is based on data of the daily incidence of COVID-19 across six large US states over the first pandemic year. We find that, on average, probabilistic forecasts from statistical time series models are overall at least as accurate as forecasts from mechanistic models. Moreover, statistical time series models better capture volatility. Our findings suggest that domain knowledge, which is integrated into mechanistic models by making assumptions about disease dynamics, does not improve short-term forecasts of disease incidence. We note, however, that forecasting is often only one of many objectives and thus mechanistic models remain important, for example, to model the impact of vaccines or the emergence of new variants. 
\par
\end{abstract}

\flushbottom
\maketitle
\thispagestyle{empty}


\newpage

\sloppy
\raggedbottom

\section{Introduction} 

Epidemic forecasting makes predictions about the development of observed quantities such as disease incidence that can inform health policy\cite{Biggerstaff2018,Meltzer2014,Bogoch2016,Nsoesie2013,Bhatia2021,Massonnaud2020}, \eg by influencing decisions about resource allocation or the timing of public health interventions to control an outbreak. Forecasts can be made over long and short time horizons with different aims\cite{Bracher2021}. Long-term forecasts (months to years) often analyze hypothetical scenarios or, for example, how an epidemic develops with increased vaccination uptake or a new, more transmissible virus strain. By contrast, short-term forecasts (days to weeks) estimate what will happen in the near future. As a result, the performance of short-term forecasts is usually evaluated by comparing them with the observed outcome. 

Different types of models can be used for epidemic forecasting, but these can be very broadly categorized into mechanistic and statistical time series models. \emph{Mechanistic models}\cite{Cori2013,Abbott2020} make assumptions about the data-generating process and are based on mathematically explicit expressions about disease dynamics. They establish a link between observed public health outcomes, such as the reported number of cases or deaths, and unobserved outcomes directly related to transmission, such as the number of infections or the instantaneous reproduction number. Formalizing these links requires disease-specific knowledge about epidemiological parameters such as the serial interval or incubation period. \emph{Statistical time series models}\cite{Hyndman2018,Taylor2018,Rasmussen2006} can be estimated without making assumptions about the data-generating process. Instead, these models detect statistical patterns in the observed time series, for example, through leveraging autocorrelation in the data. Since these patterns can be detected without domain knowledge, the statistical properties of these models are well characterized, and they can be used in a variety of forecasting applications. 

Here, we compare short-term probabilistic forecasts between mechanistic and statistical time series models. Mechanistic models provide probabilistic forecasts by taking into account uncertainty about the epidemiological process. For example, mechanistic models use prior estimates of the generation or serial interval, which are subject to uncertainty, especially during early stages of an epidemic. Statistical time series models provide probabilistic forecasts via Bayesian inference where model parameters are considered random quantities to be integrated over. A comparison of the probabilistic forecasts from mechanistic and statistical time series models can inform about the added value of domain knowledge. If forecasts from mechanistic models were more accurate, this would suggest that knowledge about disease dynamics and epidemiological parameters provide added value for short-term epidemic forecasting. 

We compare the performance of short-term probabilistic forecasts from three statistical time series models (seasonal autoregressive integrated moving average (SARIMA)\cite{Hyndman2018}, Prophet\cite{Taylor2018,Taylor2021}, Gaussian process (GP)\cite{Rasmussen2006}) with those from two popular mechanistic models based on the renewal equation (EpiEstim\cite{Cori2013,Cori2021} and EpiNow2\cite{Abbott2020,Abbott2021}). We note that there are equivalences between mechanistic models based on renewal equations and the susceptible-infected-recovered model\cite{Champredon2018-vj}. Our empirical comparison is based on data for the incidence of COVID-19 from six large US states for the first pandemic year between 15 March 2020 (the onset of the pandemic in the US) and 15 March 2021. Models are estimated repeatedly for each day using data from the last two months and, as in related work\cite{Bracher2021,Jewell2020,Krymova2022}, probabilistic forecasts are made for the following 2~weeks. Forecasting performance is evaluated retrospectively by computing the difference between model-based forecasts and observed incidence. 

\section{Materials and methods}\label{sec:methods} 

\subsection{Data}

We use daily data on the reported number of new cases of COVID-19 for six large US states (Arizona, California, Illinois, Maryland, New Jersey, and New York) from 19 January 2020 to 15 March 2021. Data are retrieved using the R package \texttt{covidcast} version~0.4.2\cite{Arnold2021}, which aggregates data on the daily number of confirmed cases of COVID-19 from the Johns Hopkins Coronavirus Resource Center\cite{Dong2020} and computes the incidence (\ie new cases per 100,000 people) using population data from the World Bank. Data are as of 1 January 2022, thus including revisions to real-time data. These revisions resulted in a negative number of new cases in a few instances, \eg when a large number of duplicated records were removed. We replaced negative values (which are implausible) with the value from the day before during data preprocessing. 

\subsection{Forecasting task}

The forecasting task is to make a daily 2-week forecast of the incidence of COVID-19 at day $t$ in US state $s = 1, \dots, 6$. Let $N_{s,t}$ denote the observed number of new cases of COVID-19 and let $\hat{N}_{s,t+1}$ to $\hat{N}_{s,t+14}$ denote the daily 2-week ahead forecasts of the number of new cases. Model-based forecasts are generated for 51~weeks (\ie 1~year) between 15 March 2020 and 15 March 2021. Forecasts are made on Sundays for the following 14~days starting on Monday. Each forecast is based on a model estimated using the historical time series up to day $t$. Specifically, the observed data two months prior to $t$ (\ie $N_{s,t-56}, \dots, N_{s,t}$) are used as training data for model $m$ to generate the probabilistic forecast $\hat{N}_{s,t+k}^{m,d}$, where $\hat{N}$ is the forecasted number of new cases and where $d = 1, \dots, D$ denotes the draw (total number of $D$ draws) from the posterior distribution $F^m_{s,t}$ of the model $m = 1, \dots, M$ (total number of $M$ models).

\subsection{Summary of forecasting models}

Table~\ref{tab:model-summary} shows a summary of the forecasting models considered in this work. The properties of each model and the implementation are described in greater detail in the following two sections. Note that we consider different modeling choices for each model by varying important model parameters or specifications. In the results, we first report for each model the modeling choice that achieved the best overall forecasting score. The models corresponding to these choices are then selected for comparison with each other.   

\begin{table}[!htpb]
    \footnotesize
    \centering
    \caption{Summary of probabilistic forecasting models and modeling choices.}
    \label{tab:model-summary}
    \begin{tabular}{p{2.5cm}p{13.5cm}}
    \midrule
    \multicolumn{2}{l}{\textbf{Mechanistic models}} \\ 
    \midrule
    (i)~EpiEstim & \tabitem Estimates the time-varying reproduction number over a fixed window $\tau$. \\
    & \tabitem Posterior distribution of reproduction number is derived analytically. \\
    & \tabitem Assumes cases by date of infection with a constant time shift. \\
    & \tabitem Uncertain generation time distribution. \\
    & $\Rightarrow$ Modeling choice: different values for $\tau$. \\
    \midrule
    (ii)~EpiNow2 & \tabitem Extension of EpiEstim. \\
    & \tabitem Different methods to model the time-varying reproduction number. \\
    & \tabitem Uncertain delay distributions for the time from infection to reporting. \\
    & \tabitem Observational model with adjustment for weekday effects. \\
    & $\Rightarrow$ Modeling choice: different methods to model the time-varying reproduction number. \\
    \midrule
    \multicolumn{2}{l}{\textbf{Statistical time series models}} \\
    \midrule
    (i)~SARIMA & \tabitem Seasonal autoregressive moving average model.  \\
    & \tabitem Weekly differencing to account for weekday effects. \\
    & $\Rightarrow$ Modeling choice: different composition of the non-seasonal part of the model. \\
    \midrule
    (ii)~Prophet & \tabitem Additive time series model. \\ 
    & \tabitem Trend component with linear growth model. \\
    & \tabitem Growth rate changes at estimated change points determined by parameter $\tau$. \\
    & \tabitem Seasonal component for weekday effects. \\
    & $\Rightarrow$ Modeling choice: different values for $\tau$. \\
    \midrule
    (iii)~GP & \tabitem Composition of kernels to model separate aspects of the time series. \\
    & \tabitem Long-range squared exponential kernel for the trend. \\
    & \tabitem Short-range squared exponential kernel for the short-term variation. \\
    & \tabitem Periodic kernel for weekday effects. \\
    & $\Rightarrow$ Modeling choice: different length of the short-range kernel. \\ 
    \midrule
    \end{tabular}
\end{table}

\subsection{Mechanistic models}

We use the following mechanistic models based on the renewal equation: EpiEstim\cite{Cori2013,Cori2021} and EpiNow2\cite{Abbott2020,Abbott2021}. 

\subsubsection*{(i)~EpiEstim} 

EpiEstim builds upon the framework by Cori et al.\cite{Cori2013} for estimating the time-varying reproduction number, which has been widely used for monitoring transmission of infectious diseases\cite{Riccardo2020,Wahaibi2020,Bhatia2021,Huisman2022}. The framework assumes that the number of new infections can be modeled with a Poisson renewal process where the number of newly infected individuals is proportional to the number of prevalent cases multiplied by their infectiousness\cite{Fraser2007}. Formally, the expected incidence on day $t$ in state $s$ is estimated as 
\begin{align}
    E\left[I_{s,t}\right] = R_{s,t} \sum_{j=0}^n w_j I_{t-j,s},    
\end{align}
where $R_{s,t}$ is the time-varying reproduction number, $w$ is the generation time (often assumed to be the serial interval distribution\cite{Flaxman2020}), and $I_{t-j,s}$ are the number of new infections in the days $j=0,\dots, n$ before $t$. Cori et al. estimate $R_{s,t}$ with a Bayesian approach and derive the posterior distribution of $R_{s,t}$ analytically by assuming a conjugate Gamma prior distribution for $R_{s,t}$. The framework by Cori et al. is implemented in the R package \texttt{EpiEstim} (version~2.2-4)\cite{Cori2021}. 

We apply EpiEstim by assuming a constant incubation period in which case the incidence of confirmed cases is exactly the incidence of infections, but shifted in time. We use an uncertain generation time distribution informed by prior estimates for the average (4.2, 95\%-interval 3.3 to 5.3) and standard deviation (4.9, 95\%-interval 3.0 to 8.3) of the generation time of SARS-CoV-2\cite{Ganyani2020}. Furthermore, we assume a Gamma prior for the basic reproduction number $R_0$ with a mean of 1.10 and standard deviation of 0.04 as in related work\cite{Bosse2021}. EpiEstim estimates the reproduction number $R_{s,t}$ over a time window $\left[t-\tau+1;t\right]$. The level of variation of the estimates for $R_{s,t}$ can be controlled by the parameter $\tau$. Larger time windows (\ie higher values for $\tau$) reduce variation in the estimates of $R_t$ and translate into changes in incidence that are more smooth. Empirically, we considered different modeling choices where we varied the time window from one day to two weeks. 

EpiEstim returns the estimated mean and standard deviation of $R_t$ in each time window. We use the mean and standard deviation from the last time window (\ie the one right before the forecast is made) to compute the shape and scale parameter for the posterior Gamma distribution of $R_{s,t}$. Based on that, we sample posterior draws  $R_{s,t}^1, \dots, R_{s,t}^D$ for each sample of the generation time distribution, which are used together with the observed incidence $N_{s,0}, \dots, N_{s,t}$ to project future incidence $\hat{N}_{s,t+1}^d$ to $\hat{N}_{s,t+21}^d$. Incidence is modeled with a Poisson distribution. Projections are made with the R package \texttt{projections} version 0.5.4\cite{Jombart2021}. 

\subsubsection*{(ii)~EpiNow2}

EpiNow2 builds upon the framework by Abbott et al.\cite{Abbott2020} to estimate the time-varying reproduction number and has been used in recent work for forecasting the incidence of COVID-19\cite{Abbott2020,Davies2021,Bracher2021}. EpiNow2 uses a similar approach to EpiEstim, yet with multiple extensions. First, uncertainty about the estimated reproduction number can be modeled using, for example, a random walk or Gaussian process. Second, the latent number of new infections can be adjusted to account for the proportion of the population that is still susceptible to the virus. Third, EpiNow2 incorporates an observation model where the expected number of new infections are mapped to the observed number of cases via uncertain delay distributions that adjust for the time from infection to reporting of a case. Fourth, the observed number of cases can be can be adjusted for weekday effects. The framework by Abbott et al. is implemented in the popular R package \texttt{EpiNow2} (version~1.3.2)\cite{Abbott2021}. 

We apply EpiNow2 by incorporating the observational model. Uncertain delay distributions are specified using prior estimates for the mean and standard deviation of the generation time\cite{Ganyani2020} (same as for EpiEstim), the incubation period\cite{Lauer2020}, and the reporting delay\cite{Cereda2020}. Furthermore, we assume a log-normal prior for the basic reproduction number $R_0$ parameterized to have the same mean and standard deviation as the Gamma distribution in EpiEstim.  Empirically, we considered different models to estimate the time-varying reproduction number, using non-parametric backcalculation, a weekly random walk, or a Gaussian process parameterized as in related work\cite{Bosse2022}. In contrast to EpiEstim, observed cases are modeled with a negative~binomial rather than a Poisson distribution, and are adjusted for weekday effects.

\subsection{Statistical time series models}

We use the following statistical time series models: a seasonal autoregressive integrated moving average (SARIMA) model\cite{Hyndman2018}, Prophet\cite{Taylor2018}, and a Gaussian process (GP)\cite{Rasmussen2006}. Different to mechanistic models, we directly model the log incidence $Y_{s,t}$ (\ie the log of the confirmed number of daily new cases per 100,000 people). Forecasts from statistical time series models are transformed to yield the number of number of new cases $N_{s,t}$ for comparison with the forecasts from mechanistic models.

\subsubsection*{(i)~SARIMA} 

SARIMA models\cite{Hyndman2018} have been used by various studies to forecast the incidence of COVID-19\cite{Alabdulrazzaq2021,Kufel2020,Wang2022,Rahman2022,Perone2022}. They are a linear combination of two components: the autoregressive (AR) and the moving average (MA) component. In the AR, the outcome $Y_{s,t}$ only depends on the lagged outcome $Y_{s,t-j}$, where $j$ denotes the time lag. In the MA, $Y_{s,t}$ only depends on the lagged errors $\epsilon_{s,t-j} = Y_{s,t-j} - Y_{s,t-j-1}$, \ie the difference between the observed and estimated outcome. 

A SARIMA($p=1$,$d=0$,$q=1$)(0,1,0)$_7$ with one AR and one MA component can be written as
\begin{align}
   \Delta Y_{s,t} = \alpha + \beta \, \Delta Y_{s,t-1} + \phi \, \Delta \epsilon_{s,t},
\end{align}
where $\Delta Y_{s,t} = Y_{s,t} - Y_{s,t-7}$ and $\Delta \epsilon_{s,t} = \epsilon_{s,t} - \epsilon_{s,t-7}$, respectively. Note that we use weekly differences of the log incidence in the seasonal part of the model to account for weekday effects. Empirically, we considered different non-seasonal parts by varying the orders of the AR ($p$), MA ($q$), and differencing ($d$). We use weakly informative priors for the model intercept ($\alpha$), parameters of the AR ($\beta$), and the MA ($\phi$)
\begin{align}
    \alpha \sim \mathrm{Normal}(0, 2.5, 6), \quad \beta \sim \mathrm{Normal}(0, 0.5), \quad \phi \sim \mathrm{Normal}(0, 0.5).
\end{align} 
The model is estimated with the R package \texttt{bayesforecast} version~1.0.1\cite{Matamoros2021}

\subsubsection*{(ii)~Prophet} 

Prophet\cite{Taylor2018} is a popular open source software in many forecasting applications and has been used by various studies to model the incidence of COVID-19\cite{Battineni2020,Satrio2021,Wang2022}. It is an additive model with three components modeling the trend of the time series, seasonality, and holiday effects (not considered here). The crucial component is the trend. It can be modeled assuming linear or logistic growth and the growth rate is allowed to change at times $s_j$, $j = 1,\dots,J$. The changes in growth rates $\delta_j$ are modeled with a hierarchical prior $\delta_j \sim \mathrm{Laplace}(0,\tau)$. The number of times the growth rate can change can be controlled by adjusting the scale of the prior ($\tau$). Greater values of $\tau$ translate into more frequent changes of the trend. 

We use Prophet assuming linear growth for the trend component and account for weekday effects in the seasonality component. Empirically, we considered different values for the scale of the prior for the number of change points in the range of $\tau \in \left[0.10, 0.45\right]$. The implementation is based on the R package \texttt{prophet} version~1.0\cite{Taylor2021}.

\subsubsection*{(iii)~Gaussian process (GP)} 

GPs\cite{Rasmussen2006} have been used in epidemic forecasting\cite{Johnson2018,Velasquez2020,Qian2020}. A GP is a stochastic process or collection of random variables such that the joint distribution of every finite subset is multivariate normal (Gaussian) distributed. A GP is completely defined by a function for the mean and covariance (commonly called the kernel function). The mean function can be specified arbitrarily but is often set to zero after the data is transformed, \eg centered. The crucial component is therefore the kernel, which characterises the correlation between pairs of input data. Prior information is reflected in the choice of the kernel used to describe the behavior of the stochastic process. A popular kernel is the squared exponential (SE) kernel $k_{\mathrm{SE}}$ defined as
\begin{align}
    k_{\mathrm{SE}}(t_i,t_j) = \alpha^2 \exp\left(- \frac{\left(t_i - t_j\right)^2}{2\rho^2}\right),
\end{align}
where $\alpha^2$ is the signal noise parameter determining the scale of the output and $\rho$ is the length scale (or bandwidth) parameter determining the correlation between input pairs, \ie two time points $(t_i,t_j)$. Larger $\alpha$ allow for greater deviations from the mean and larger $\rho$ result in functions with rougher paths. Often one of the two parameters $\rho$ and $\alpha$ is fixed due to weak identifiability\cite{Rasmussen2006,Hawyrluk2021}. For time series analysis, seasonal patterns can be modeled using a periodic kernel (PE)
\begin{align}
    k_{\mathrm{PE}}(t_i,t_j) = \alpha^2 \exp\left(- \frac{2\,\sin\left(\pi\,|t_i - t_j|\,/\,p\right)^2}{2\rho^2}\right),
\end{align}
where $p$ is an additional parameter for the periodicity.

We model $Y_{s,t}$ with a zero mean GP using a composition of three kernels
\begin{align}
    Y_{s,t} &\sim \mathrm{GP}\left(0,k_{\mathrm{SE}}^{\mathrm{long}} + k_{\mathrm{SE}}^{\mathrm{short}} + k_{\mathrm{PE}}^{\mathrm{week}} + \sigma^2 \delta_{ij}\right)~,
\end{align}
where $k_{\mathrm{SE}}^{\mathrm{long}}$ is a SE kernel with a long length scale and $k_{\mathrm{SE}}^{\mathrm{short}}$ is SE kernel with short length scale, $k_{\mathrm{PE}}^{\mathrm{week}}$ is PE kernel for weekday effects, and $\sigma^2 \delta_{ij}$ is a regularizing term with a Kronecker delta function ensuring $\sigma^2$ Gaussian noise is only added when $i=j$. The SE kernels model the long-term trend and the short-term variation using a long- and short-range correlation structure\cite{Hawyrluk2021}. 

Prior specifications for each kernel are as follows. For $k_{\mathrm{SE}}^{\mathrm{long}}$, we set the length scale to $\rho = 56$ days (the length of the training period) and use a weakly informative prior for $\alpha \sim \text{Student-t}(\nu=3,\mu=0,\sigma=5)$. For $k_{\mathrm{PE}}^{\mathrm{week}}$, we set the length scale to $\rho = 1$, the periodicity to $p = 7$ days, and use $\alpha \sim \text{Student-t}(\nu=5,\mu=0,\sigma=2)$. Empirically, we considered different length scales $\rho \in (1,7,14)$ days in the short-range kernel $k_{\mathrm{SE}}^{\mathrm{short}}$, allowing the GP to react more or less quickly to short-term variation. A weakly informative prior is given by $\alpha \sim \text{Student-t}(\nu=7,\mu=0,\sigma=2)$. We implemented the GPs in the probabilistic programming language Stan version~2.21.0\cite{Carpenter2017}.

\subsection{Evaluation measures}

We evaluate probabilistic forecasts over a 2-week time window. We aggregate daily forecasts by week to eliminate potential weekday effects, which is not accounted for in the EpiEstim model. The following evaluation measures are separately computed for the 1- and 2-week ahead forecast. 

Forecasting performance is primarily assessed based on the continuous ranked probability score (CRPS)\cite{Gneiting2007}. It measures the difference between the forecasted and the observed value and is a strictly proper scoring rule. It is defined as 
\begin{align}
    CRPS(F_f,x_t)= \int \big[F(y_t) - \mathbf{1}(y_t \geq x_t)\big]^2 \,\mathrm{d}y_t,
\end{align}
where $F$ is the cumulative distribution function of the forecast $y$ for observation $x$ at time $t$. Note that the CRPS is a generalization of the absolute error between the forecasted and observed value. A lower CRPS therefore indicates a more accurate forecast. 

We follow recommendations to compute CRPS for the log of the forecasted incidence\cite{Bosse2023}. This corresponds to evaluating the multiplicative rather than the absolute forecasting error. Importantly, taking the log mitigates the influence of extreme outliers in the forecasts often made by models during an exponentially growing epidemic.

CRPS is computed for each forecast and scores are summarized by model, state, and epidemic phase (labeled manually, see \supp~Fig.~\zref{fig:phases}). In addition, we perform pairwise model comparisons\cite{Bosse2022_Rpackage}, thereby measuring the forecasting performance of each model relative to others. This is accomplished by first computing the ratio of scores. The relative skill is then computed by taking the geometric mean of all score ratios. Lower score ratios (and lower relative skill) indicate better forecasting performance in pairwise comparisons. 

Secondary performance measures include sharpness, calibration, and bias. Sharpness is measured as the normalized median of the absolute deviation from the median of the predictive samples\cite{Funk2019}. Calibration is assessed based on the density of the probability integral transform (PIT) values\cite{Gneiting2007} as well as the empirical coverage of the 50\,\% and 95\,\% predictive interval (PI)\cite{Bracher2021}. Bias is evaluated with the average proportion of posterior draws that are above the observed value.

Primary and secondary performance measures are computed using the R packages \texttt{scoringRules} version~1.0.2\cite{Jordan2022} and \texttt{scoringutils}\cite{Bosse2022_Rpackage} version~1.0.1.

An important quality of epidemic forecasts is to anticipate exponential growth. This quality is not captured specifically by the CRPS. Therefore, we consider an additional metric by McDonald et al.\cite{McDonald2021} for detecting so-called ``hotspots''. A hotspot at time $t$ in state $s$ is defined as a weekly increase in incidence of more than 25\%, \ie
\begin{align}
    H_{s,t} = 
    \begin{cases}
1, \quad \text{if }~(I_{s,t}-I_{s,t-7})/(I_{s,t-7}) \geq 0.25,\\
0, \quad \text{otherwise.}
\end{cases}
\end{align}
The probability for a hotspot predicted by model $m$ is then equal to the proportion of posterior draws where the weekly change in the forecast is larger than 25\%. Performance is assessed with the area under the receiving operating characteristic curve (AUC). Similar to McDonald et al.\cite{McDonald2021}, we exclude data where the weekly incidence is below 70~new cases per 100,000 people. This removes 35\% of the observations. Of the remaining observations, 25\% are classified as hotspots. 

\subsection{Software and model estimation}

All analyses are performed in the software R (version 4.2.0)\cite{RCoreTeam2022}. Probabilistic models are estimated using Markov chain Monte Carlo (MCMC) sampling (except for EpiEstim where the posterior is derived analytical). Each model is estimated with 4~Markov chains and 1,000~iterations of which the first 500~iterations were discarded as part of the warm-up. This results in 2,000 posterior samples. 

\subsection{Data and code availability}

Data are from publicly available data sources (Johns Hopkins Coronavirus Resource Center for epidemiological data; World Bank for population data). Preprocessed data files together with reproducible code is available from \url{https://github.com/nbanho/covid_predict}.

\FloatBarrier

\section{Results}\label{sec:results}

The specific modeling choice has only a marginal influence on the overall forecasting performance of each model (\supp~Fig.~\zref{fig:experiments}). Table~\ref{tab:best-specifications} shows the modeling choice with the highest overall CRPS for each model. The models corresponding to these choices are used for the subsequent between-model comparisons of mechanistic and statistical time series models. 

\begin{table}[!htpb]
    \footnotesize
    \centering
    \caption{\textbf{Modeling choices with highest overall CRPS.} Results for all choices are shown in \supp~Fig.~\zref{fig:experiments}.}
    \label{tab:best-specifications}
    \begin{tabular}{ll}
    \toprule
    Model & Modeling choice with highest overall CRPS \\
    \midrule
    \multicolumn{2}{l}{\textbf{Mechanistic models}} \\ 
    \midrule
    EpiEstim & Time-varying reproduction number estimated over a 7-day time window. \\ 
    EpiNow2 & Time-varying reproduction number modeled with a Gaussian process. \\
    \midrule
    \multicolumn{2}{l}{\textbf{Statistical time series models}} \\
    \midrule
    SARIMA & SARIMA(1,0,1)(0,1,0)$_7$ model with one AR and one MA component. \\
    Prophet & Prior scale $\tau = 0.45$ allowing for high number of change points for the growth rate. \\
    GP & Short-range kernel with a length scale of $\rho = 7\,$days. \\
    \bottomrule
    \end{tabular}
\end{table}

\begin{figure}[!htpb]
    \centering
    \includegraphics[width=\linewidth]{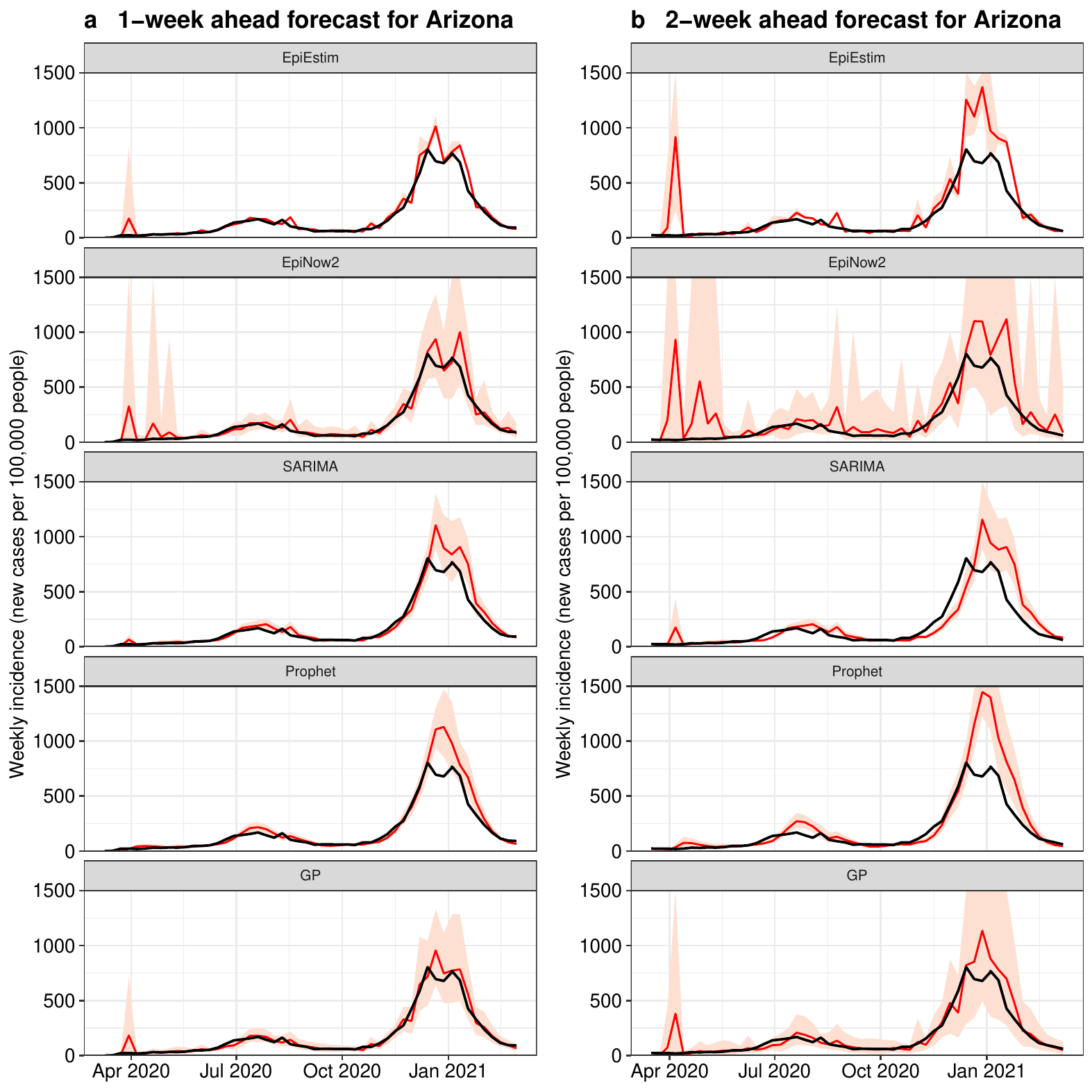}
    \caption{\textbf{Probabilistic forecasts for Arizona by model.} Probabilistic forecast (posterior mean as red line, 95\%-PI as shaded area) for Arizona by each model. \textbf{(a)}~1-week ahead forecast. \textbf{(b)}~2-week ahead forecast. Observed incidence is shown with black lines. Probabilistic forecasts for all US states by model are shown in \supp~\zref{fig:forecasts-az}-\zref{fig:forecasts-ny}.}
    \label{fig:example}
\end{figure}

The 1- and 2-week ahead forecasts by each model for Arizona are shown as examples in Fig.~\ref{fig:example}. The 1-week ahead forecast is well-calibrated across all models and covers the observed weekly incidence most of the time, except for the peak in January 2021 for SARIMA and Prophet. Coverage becomes worse for the 2-week ahead forecast, especially the peak in January 2021 is overestimated by most models and the subsequent decline is not anticipated. Forecasts from EpiNow2 are wider than those from EpiEstim. Similarly, forecasts from GP are wider than those from SARIMA and Prophet.  

\begin{figure}[!htpb]
    \centering
    \includegraphics[width=\linewidth]{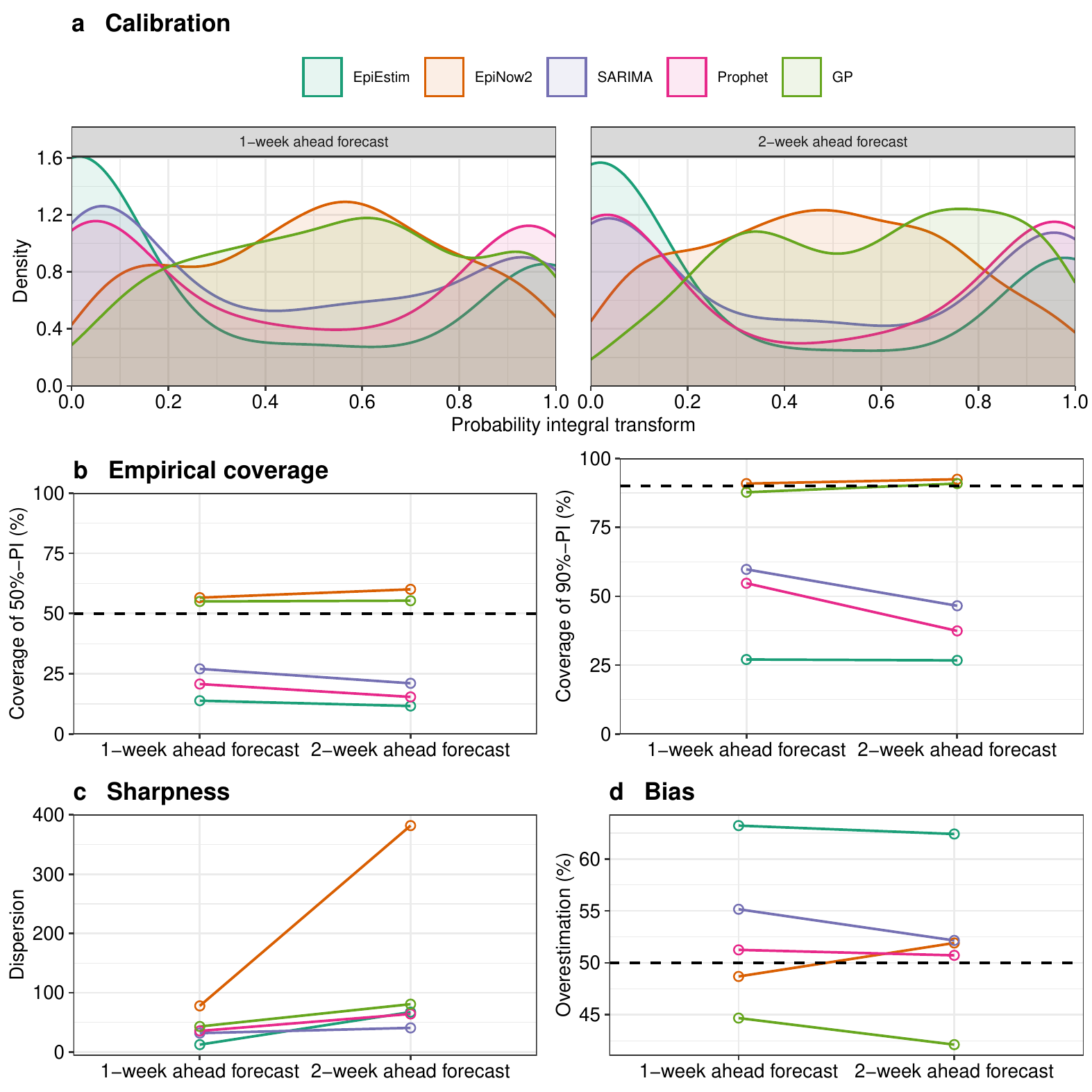}
    \caption{\textbf{Evaluation of secondary performance measures overall.} \textbf{(a)}~Probabilistic calibration assessed with the density distribution of the probability integral transform. \textbf{(b-c)}~Coverage of the 50\,\% (left panel) and 95\,\% (right panel) predictive interval [PI]. \textbf{(d)}~Sharpness assessed with dispersion measured median of the absolute deviation from the median of the predictive samples. \textbf{(e)}~Bias assessed with the average proportion of posterior draws above observed the incidence.}
    \label{fig:descriptive}
\end{figure}

Results for secondary performance measures (calibration, sharpness, and bias) across states are shown in Fig.~\ref{fig:descriptive}. The distribution of the probability integral transform for EpiEstim, SARIMA, and Prophet is U-shaped (Fig.~\ref{fig:descriptive}a), indicating that the predictive distribution is too narrow, whereas it is hump-shaped for EpiNow2 and GP, indicating that the predictive distribution is too wide. GP and EpiNow2 achieve good coverage of the 50\%-PI and the 95\%-PI, whereas EpiEstim, SARIMA, and Prophet cover both PIs less often than theoretically predicted (Fig.~\ref{fig:descriptive}b). Forecasts from EpiNow2 are least sharp, especially the 2-week ahead forecasts exhibit by far the highest dispersion (Fig.~\ref{fig:descriptive}c). EpiEstim tends to overestimate and GP tends to underestimate the true incidence (Fig.~\ref{fig:descriptive}d). Other models do not show a similar behavior in terms of bias.

\begin{figure}[!htpb]
    \centering
    \includegraphics{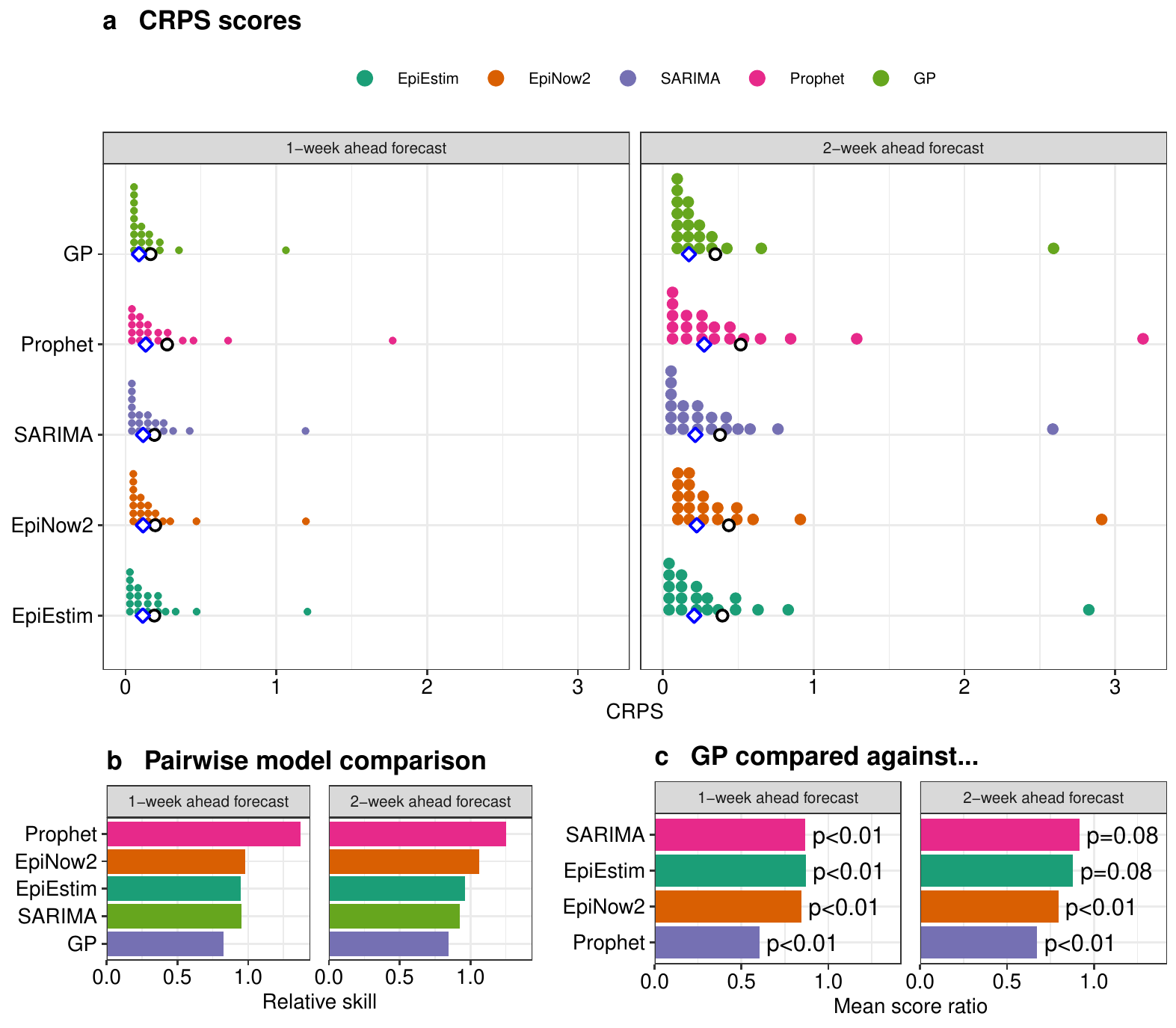}
    \caption{\textbf{Evaluation of forecasting performance.} \textbf{(a)}~Empirical quantile distribution of continuous ranked probability score (CRPS) by model. Each dot represents one of the 20~quantiles of the empirical distribution. Median CRPS is shown with the blue square and mean CRPS with the black dot. Lower CRPS indicate better forecasting performance. \textbf{(b)}~Relative skill for pairwise model comparisons according to CRPS scores. Lower values indicate better relative forecasting performance. \textbf{(c)}~Mean score ratio for GP compared against other models. Score ratios closer to 1 indicate forecasting performances more similar to GP. }
    \label{fig:crps_by_week}
\end{figure}

CRPS scores are shown in Fig.~\ref{fig:crps_by_week}. The distribution of scores are right-skewed for all models, indicating a considerable number of large forecasting errors (Fig.~\ref{fig:crps_by_week}a). For the 1-week ahead forecast, the scores are comparable between models. Mean and median CRPS are lowest (best) for GP and highest for Prophet. The latter also has the highest CRPS for the 95\% quantile. CRPS scores are more dispersed for the 2-week ahead forecasts, but the differences between models remain similar, albeit more visible. GP achieves the best scores, and the distribution is more concentrated on low CRPS.  By contrast, the distribution is more dispersed for Prophet and the CRPS for the 90\% and 95\% quantiles are higher. CRPS scores for SARIMA, EpiNow2, and EpiEstim are comparable and in between those of GP and Prophet. Differences in forecasting performance are confirmed by pairwise model comparisons. GP has the lowest (best) relative skill for both the 1- and 2-week ahead forecast (Fig.~\ref{fig:crps_by_week}b) . Compared against other models, GP achieves the highest (best) score ratio, although the difference to SARIMA and EpiEstim are statistically not significant for the 2-week ahead forecast (Fig.~\ref{fig:crps_by_week}c). In addition, GP has the lowest CRPS scores in all six US states (\supp~Fig.~\zref{fig:crps_by_state}) and performs particularly well in phases of (sub)exponential decline (\supp~Fig.~\zref{fig:crps_by_phase}) because it anticipates sooner the peak and subsequent decline of incidence.  

\begin{figure}[!htpb]
    \centering
    \includegraphics{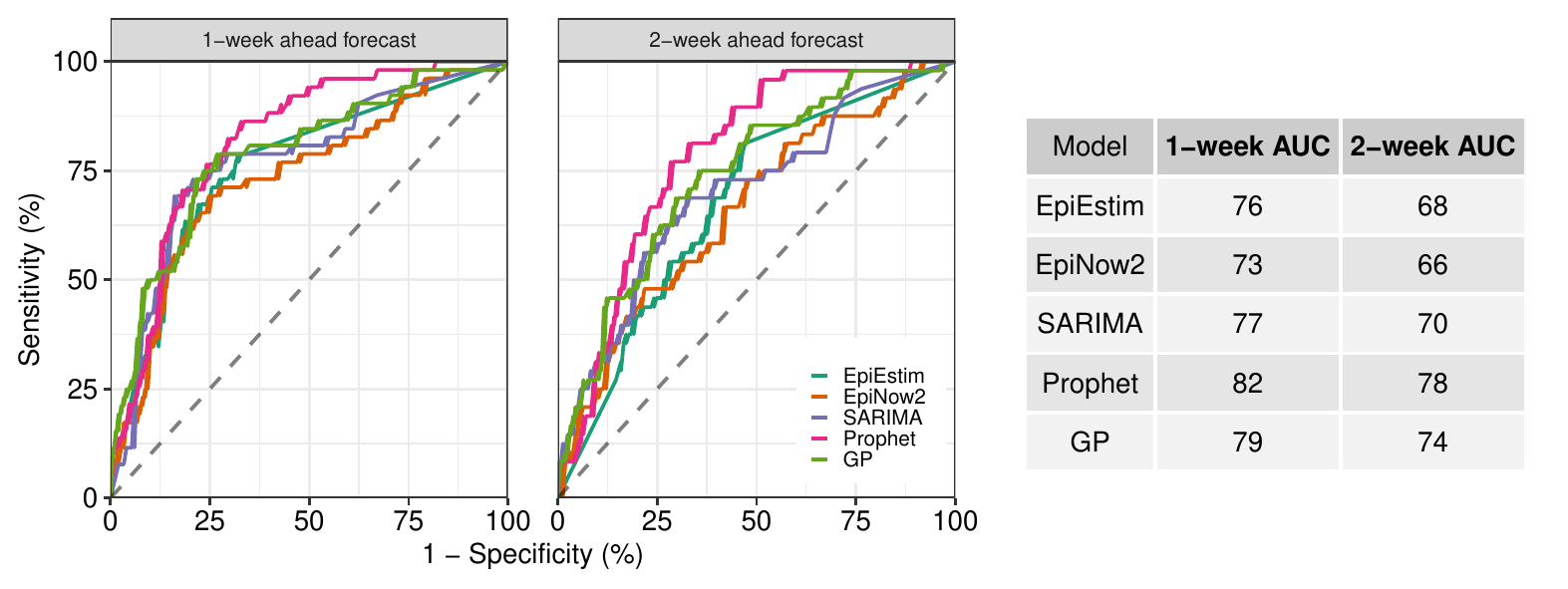}
    \caption{\textbf{Evaluation of hotspot prediction.} Area under the receiver operating characteristic curve (AUC) for predicting hotspot, defined as a weekly increase in incidence of more than 25\% to more than 10 new cases per 100,000 people). Higher AUC indicate greater ability to detect hotspots.}
    \label{fig:hotspot}
\end{figure}

The CRPS evaluates the overall forecasting performance. Another important quality of forecasting models is to predict hotspots, \ie large increases in weekly incidence. A comparison of the AUC for predicting hotspots shows that statistical time series models overall predict hotspots more accurately than mechanistic models (Fig.~\ref{fig:hotspot}). Prophet has the highest AUC  for both the 1- and 2-week ahead forecast, followed by GP and SARIMA.  

\FloatBarrier

\section{Discussion}\label{sec:discussion}

\subsection{Summary}

We compared the short-term forecasts from mechanistic and statistical time series models for the incidence of COVID-19 in six large US states during the first year of pandemic. Forecasts from statistical time series models were at least as accurate as those from mechanistic models. GP achieved the best overall forecasting performance, but the difference to other models such as SARIMA and EpiEstim were not large, especially for the 2-week ahead forecast. The forecasts from GP and EpiNow2 were less sharp than those from other models (EpiEstim, SARIMA, and Prophet), but their empirical coverage was better. Finally, statistical time series models predicted large increases in incidence more accurately than mechanistic models.

\subsection{Findings in context}

Our work shows that statistical time series models achieve similar if not better short-term forecasting performance than mechanistic models. This is in line with previous work showing that statistical time series models with various known applications can also provide reasonably accurate forecasts for disease spread\cite{Wang2022,Rahman2022,Johnson2018}. Nevertheless, epidemic forecasts remain difficult for both types of models. The reason is that forecasts are better understood as projections that are entirely based on the historical time series of the outcome. As such, these projections assume that nothing is changing and that the historical trend continues into the future. However, there can be factors influencing disease spread that are not captured by the historical time series, \eg voluntary behavioral changes or non-pharmaceutical interventions\cite{Brauner2021, Banholzer2021,Flaxman2020}.

A related study compared human- and model-based forecasts for the cases and deaths of COVID-19 up to four weeks ahead\cite{Bosse2022}. Human-based forecasts were elicited using a web application where participants could select a predictive distribution (\eg log-normal), and then visualize and interactively adjust their numerical forecasts. Individual forecasts were aggregated with a quantile-wise mean and the crowd forecasts were compared to the probabilistic forecasts from mechanistic models. The study showed that crowd forecasts were more accurate than forecasts from mechanistic models. In particular, mechanistic models often overestimated the incidence at the epidemic peaks. Similarly, we found in our study that the peak incidence was often overestimated by mechanistic models, but also statistical time series models. An explanation for why model-based forecasts overestimate the peak compared to human-based forecasts may be that only the latter can factor in the possible impact of non-pharmaceutical interventions that are implemented right at the time when forecasts have to be made.

Model-based forecasts of disease incidence could be further improved by using ensembles or including auxiliary indicators. Previous studies showed that the combined forecast of multiple models can be more accurate than the forecasts from each individual model\cite{Bracher2021,Cramer2022,Sherratt2022}, and that additional data streams (\eg online surveys or mobility) can improve forecasting performance\cite{McDonald2021,Reinhart2021,Schwabe2021,Rodriguez2021}. Ensemble forecasts and auxiliary indicators were not the focus of this work, but we acknowledge their potential. 

\subsection{Limitations}

Our study has limitations. First, we evaluated the multiplicative error of forecasts by comparing the log incidence. We chose the multiplicative error as it is less sensitive to outliers and ensures comparability across states and epidemic phases\cite{Bosse2022}. A drawback of the multiplicative error is that it is the same regardless of the absolute level of incidence. This can be unfavorable to models with large errors during phases of low incidence, which are less concerning. Second, we used the CRPS as our primary performance measure to evaluate forecasts. The CRPS is a proper scoring rule that is used across multiple domains to evaluate numerical forecasts\cite{Gneiting2007}. However, policymakers may evaluate forecasts differently using a utility function which, for example, penalizes underestimation of disease incidence more than overestimation. Third, although we compared models that are commonly used to forecast incidence, they still only represent a subset of all possible models. We therefore acknowledge that superior performance may be achieved with extensions of the models considered in this work or other types of models that are not considered in this study, \eg compartmental models\cite{Reiner2021}, gradient boosted trees\cite{Rahman2022}, or deep learning\cite{Rodriguez2021}. Fourth, forecasts were made and analyzed retrospectively, which may affect their evaluation. By contrast, forecasts made in real-time are often based on incomplete incidence data\cite{McDonald2021} or forecasters may make manual adjustments before submission\cite{Bracher2021}. Fifth, we evaluated our forecasts based on the reported incidence of COVID-19. We therefore acknowledge that the actual number of cases of COVID-19 may be higher than reported number of cases\cite{Noh2021}, \ie ascertainment bias is not addressed during our analysis. Sixth, we evaluated forecasts on the incidence of COVID-19 in six large US states over one year with carefully labeled epidemic phases. Future research may extend our analysis, \eg considering different geographical regions, other epidemiological outcomes (\eg deaths\cite{Bracher2021,Cramer2022} or hospitalizations\cite{Perone2022}), and or even epidemics of other pathogens (\eg influenza\cite{Biggerstaff2018,Bogoch2016}).     

\subsection{Concluding remarks and outlook}\label{sec:conclusions}

Mechanistic models incorporate knowledge about disease dynamics. Such domain knowledge is important, for example, when modeling the effects of vaccination uptake or the emergence of new virus variants. However, the added value for short-term forecasts of disease incidence is less clear. We empirically compared short-term probabilistic forecasts of disease from mechanistic and statistical time series models. We found that statistical time series models achieved similar overall performance than mechanistic models, suggesting that disease incidence can be predicted in the short term without modeling disease dynamics. Future research may evaluate whether forecasts can be improved further by other means such as considering ensemble models or auxiliary indicators. 

\newpage

\bibliography{references.bib}

\begin{thebibliography}{10}

\bibitem{Biggerstaff2018}
Biggerstaff M, Johansson M, Alper D, Brooks LC, Chakraborty P, Farrow DC,
  et~al.
\newblock Results from the second year of a collaborative effort to forecast
  influenza seasons in the {United} {States}.
\newblock Epidemics. 2018;24:26--33.
\newblock doi:10.1016/j.epidem.2018.02.003.

\bibitem{Meltzer2014}
Meltzer MI, Atkins CY, Santibanez S, Knust B, Petersen BW, Ervin ED, et~al.
\newblock Estimating the future number of cases in the {Ebola} epidemic --
  {Liberia} and {Sierra} {Leone}, 2014–2015.
\newblock Morbidity and Mortality Weekly Report. 2014;63(3).
\newblock Available from: \url{https://stacks.cdc.gov/view/cdc/24901}.

\bibitem{Bogoch2016}
Bogoch II, Brady OJ, Kraemer MUG, German M, Creatore MI, Kulkarni MA, et~al.
\newblock Anticipating the international spread of {Zika} virus from {Brazil}.
\newblock Lancet. 2016;387(10016):335--336.
\newblock doi:10.1016/S0140-6736(16)00080-5.

\bibitem{Nsoesie2013}
Nsoesie E, Mararthe M, Brownstein J.
\newblock Forecasting {Peaks} of {Seasonal} {Influenza} {Epidemics}.
\newblock PLoS Curr. 2013;5.
\newblock doi:10.1371/currents.outbreaks.bb1e879a23137022ea79a8c508b030bc.

\bibitem{Bhatia2021}
Bhatia S, Parag KV, Wardle J, Imai N, Van~Elsland SL, Lassmann B, et~al..
  Global predictions of short- to medium-term {COVID}-19 transmission trends:
  {A} retrospective assessmen.
\newblock medRxiv [Preprint]; 2021 [cited 2022 March 3].
\newblock Available from:
  \url{https://www.medrxiv.org/content/10.1101/2021.07.19.21260746.abstract}.
  doi:10.1101/2021.07.19.21260746.

\bibitem{Massonnaud2020}
Massonnaud C, Roux J, Crépey P. {COVID}-19: {Forecasting} short term hospital
  needs in {France}.
\newblock medRxiv [Preprint]; 2020 [cited 2022 March 3].
\newblock Available from:
  \url{https://www.medrxiv.org/content/10.1101/2020.03.16.20036939v1}.
  doi:10.1101/2020.03.16.20036939.

\bibitem{Bracher2021}
Bracher J, Wolffram D, Deuschel J, Görgen K, Ketterer JL, Ullrich A, et~al.
\newblock A pre-registered short-term forecasting study of {COVID}-19 in
  {Germany} and {Poland} during the second wave.
\newblock Nat Commun. 2021;12:5173.
\newblock doi:10.1038/s41467-021-25207-0.

\bibitem{Cori2013}
Cori A, Ferguson NM, Fraser C, Cauchemez S.
\newblock A new framework and software to estimate time-varying reproduction
  numbers during epidemics.
\newblock Am J Epidemiol. 2013;178(9):1505--1512.
\newblock doi:10.1093/aje/kwt133.

\bibitem{Abbott2020}
Abbott S, Hellewell J, Thompson RN, Sherratt K, Gibbs HP, Bosse NI, et~al.
\newblock Estimating the time-varying reproduction number of {SARS}-{CoV}-2
  using national and subnational case counts.
\newblock Welcome Open Research. 2020;(5):112.
\newblock doi:10.12688/wellcomeopenres.16006.2.

\bibitem{Hyndman2018}
Hyndman RJ, Athanasopoulos G.
\newblock Chapter 8: {ARIMA} models.
\newblock In: Forecasting: principles and practice. OTexts; 2018. p. 221--274.

\bibitem{Taylor2018}
Taylor SJ, Letham B.
\newblock Forecasting at {Scale}.
\newblock Am Stat. 2018;72(1):37--45.
\newblock doi:10.1080/00031305.2017.1380080.

\bibitem{Rasmussen2006}
Rasmussen CE.
\newblock Gaussian {Processes} for {Machine} {Learning}.
\newblock MIT Press; 2006.

\bibitem{Taylor2021}
Taylor S, Letham B. Prophet: {Automatic} {Forecasting} {Procedure}; 2021.
\newblock R package version 1.0.
\newblock Available from:
  \url{https://cran.r-project.org/web/packages/prophet}.

\bibitem{Cori2021}
Cori A, Kamvar Z, Stockwin J, Jombart T, Dahlqwist E, FitzJohn R, et~al..
  {EpiEstim}: {A} tool to estimate time varying instantaneous reproduction
  number during epidemics; 2021.
\newblock R package version 2.2-4.
\newblock Available from:
  \url{https://cran.r-project.org/web/packages/EpiEstim}.

\bibitem{Abbott2021}
Abbott S, Hellewell J, Sherrat K, Gostic KM, Hickson J, Badr HS, et~al..
  {EpiNow2}: {Estimate} real-time case counts and time-varying epidemiological
  parameters; 2021.
\newblock R package version 1.3.2.
\newblock Available from:
  \url{https://cran.r-project.org/web/packages/EpiNow2}.

\bibitem{Champredon2018-vj}
Champredon D, Dushoff J, Earn DJD.
\newblock Equivalence of the {Erlang-distributed} {SEIR} epidemic model and the
  renewal equation.
\newblock SIAM J Appl Math. 2018;78(6):3258--3278.

\bibitem{Jewell2020}
Jewell NP, Lewnard JA, Jewell BL.
\newblock Predictive mathematical models of the {COVID}-19 pandemic:
  {Underlying} principles and value of projections.
\newblock JAMA. 2020;323(19):1893--1894.
\newblock doi:10.1001/jama.2020.6585.

\bibitem{Krymova2022}
Krymova E, Béjar B, Thanou D, Sun T, Manetti E, Lee G, et~al.
\newblock Trend estimation and short-term forecasting of {COVID}-19 cases and
  deaths worldwide.
\newblock Proc Natl Acad Sci U S A. 2022;119(32):e2112656119.
\newblock doi:10.1073/pnas.2112656119.

\bibitem{Arnold2021}
Arnold T, Bien J, Brooks L, Colquhoun S, Farrow D, Grabman J, et~al..
  Covidcast: {Client} for delphi's {COVIDcast} epidata {API}; 2021.
\newblock R package version 0.4.2.
\newblock Available from:
  \url{https://cran.r-project.org/web/packages/covidcast}.

\bibitem{Dong2020}
Dong E, Du H, Gardner L.
\newblock An interactive web-based dashboard to track {COVID}-19 in real time.
\newblock Lancet Infect Dis. 2020;20(5):533--534.
\newblock doi:10.1016/S1473-3099(20)30120-1.

\bibitem{Riccardo2020}
Riccardo F, Ajelli M, Andrianou XD, Bella A, Manso MD, Fabiani M, et~al.
\newblock Epidemiological characteristics of {COVID}-19 cases and estimates of
  the reproductive numbers 1 month into the epidemic, {Italy}, 28 {January} to
  31 {March} 2020.
\newblock Euro Surveill. 2020;25(49):2000790.
\newblock doi:10.2807/1560-7917.ES.2020.25.49.2000790.

\bibitem{Wahaibi2020}
Al~Wahaibi A, Al~Manji A, Al~Maani A, Al~Rawahi B, Al~Harthy K, Alyaquobi F,
  et~al.
\newblock {COVID}-19 epidemic monitoring after non-pharmaceutical
  interventions: {The} use of time-varying reproduction number in a country
  with a large migrant population.
\newblock Int J Infect Dis. 2020;99:466--472.
\newblock doi:10.1016/j.ijid.2020.08.039.

\bibitem{Huisman2022}
Huisman JS, Scire J, Angst DC, Li J, Neher RA, Maathuis MH, et~al.
\newblock Estimation and worldwide monitoring of the effective reproductive
  number of {SARS}-{CoV}-2.
\newblock eLife. 2022;11:e71345.
\newblock doi:10.7554/eLife.71345.

\bibitem{Fraser2007}
Fraser C.
\newblock Estimating individual and household reproduction numbers in an
  emerging epidemic.
\newblock PLoS One. 2007;2(8):e758.
\newblock doi:10.1371/journal.pone.0000758.

\bibitem{Flaxman2020}
Flaxman S, Mishra S, Gandy A, Unwin HJT, Mellan TA, Coupland H, et~al.
\newblock Estimating the effects of non-pharmaceutical interventions on
  {COVID}-19 in {Europe}.
\newblock Nature. 2020;584(7820):257--261.
\newblock doi:10.1038/s41586-020-2405-7.

\bibitem{Ganyani2020}
Ganyani T, Kremer C, Chen D, Torneri A, Faes C, Wallinga J, et~al.
\newblock Estimating the generation interval for coronavirus disease
  ({COVID}-19) based on symptom onset data, {March} 2020.
\newblock Euro Surveill. 2020;25(17):2000257.
\newblock doi:10.2807/1560-7917.ES.2020.25.17.2000257.

\bibitem{Bosse2021}
Bosse NI, Abbott S, Bracher J, Hain H, Quilty BJ, Jit M, et~al.. Comparing
  human and model-based forecasts of {COVID}-19 in {Germany} and {Poland}.
\newblock medRxiv [Preprint]; 2021 [cited 2022 March 3].
\newblock Available from:
  \url{https://www.medrxiv.org/content/10.1101/2021.12.01.21266598v1}.
  doi:10.1101/2021.12.01.21266598.

\bibitem{Jombart2021}
Jombart T, Nouvellet P, Bhatia S, Kamvar ZN, Taylor T, Ghozzi S. Projections:
  {Project} {Future} {Case} incidence; 2021.
\newblock R package version 0.5.4.
\newblock Available from:
  \url{https://cloud.r-project.org/web/packages/projections}.

\bibitem{Davies2021}
Davies NG, Abbott S, Barnard RC, Jarvis CI, Kucharski AJ, Munday JD, et~al.
\newblock Estimated transmissibility and impact of {SARS}-{CoV}-2 lineage
  {B}.1.1.7 in {England}.
\newblock Science. 2021;372(6538):eabg3055.
\newblock doi:10.1126/science.abg3055.

\bibitem{Lauer2020}
Lauer SA, Grantz KH, Bi Q, Jones FK, Zheng Q, Meredith HR, et~al.
\newblock The incubation period of {Coronavirus} {Disease} 2019 ({COVID}-19)
  from publicly reported confirmed cases: {Estimation} and application.
\newblock Ann Intern Med. 2020;172(9):577--582.
\newblock doi:10.7326/M20-0504.

\bibitem{Cereda2020}
Cereda D, Tirani M, Rovida F, Demicheli V, Ajelli M, Poletti P, et~al.. The
  early phase of the {COVID}-19 outbreak in {Lombardy}, {Italy}.
\newblock arXiv [Preprint]; 2020 [cited 2022 March 3].
\newblock Available from: \url{https://arxiv.org/abs/2003.09320}.
  doi:10.48550/ARXIV.2003.09320.

\bibitem{Bosse2022}
Bosse NI, Abbott S, Bracher J, Hain H, Quilty BJ, Jit M, et~al.
\newblock Comparing human and model-based forecasts of {COVID}-19 in {Germany}
  and {Poland}.
\newblock PLoS Comput Biol. 2022;18(9):e1010405.
\newblock doi:10.1371/journal.pcbi.1010405.

\bibitem{Alabdulrazzaq2021}
Alabdulrazzaq H, Alenezi MN, Rawajfih Y, Alghannam BA, Al-Hassan AA, Al-Anzi
  FS.
\newblock On the accuracy of {ARIMA} based prediction of {COVID}-19 spread.
\newblock Results Phys. 2021;27:104509.
\newblock doi:10.1016/j.rinp.2021.104509.

\bibitem{Kufel2020}
Kufel T.
\newblock {ARIMA}-based forecasting of the dynamics of confirmed {COVID}-19
  cases for selected {European} countries.
\newblock Equilibrium. 2020;15(2):181--204.
\newblock doi:10.24136/eq.2020.009.

\bibitem{Wang2022}
Wang Y, Yan Z, Wang D, Yang M, Li Z, Gong X, et~al.
\newblock Prediction and analysis of {COVID}-19 daily new cases and cumulative
  cases: times series forecasting and machine learning models.
\newblock BMC Infect Dis. 2022;22(1):495.
\newblock doi:10.1186/s12879-022-07472-6.

\bibitem{Rahman2022}
Rahman MS, Chowdhury AH, Amrin M.
\newblock Accuracy comparison of {ARIMA} and {XGBoost} forecasting models in
  predicting the incidence of {COVID}-19 in {Bangladesh}.
\newblock PLoS Global Public Health. 2022;2(5):e0000495.
\newblock doi:10.1371/journal.pgph.0000495.

\bibitem{Perone2022}
Perone G.
\newblock Comparison of {ARIMA}, {ETS}, {NNAR}, {TBATS} and hybrid models to
  forecast the second wave of {COVID}-19 hospitalizations in {Italy}.
\newblock Eur J Health Econ. 2022;23(6):917--940.
\newblock doi:10.1007/s10198-021-01347-4.

\bibitem{Matamoros2021}
Matamoros AA, Torres CC, Dala A, Hyndman R, O'Hara-Wild M. Bayesforecast:
  {Bayesian} {Time} {Series} {Modeling} with {Stan}; 2021.
\newblock R package version 1.0.1.
\newblock Available from:
  \url{https://cran.r-project.org/web/packages/bayesforecast}.

\bibitem{Battineni2020}
Battineni G, Chintalapudi N, Amenta F.
\newblock Forecasting of {COVID}-19 epidemic size in four high hitting nations
  ({USA}, {Brazil}, {India} and {Russia}) by {Fb}-{Prophet} machine learning
  model.
\newblock Appl Comput Inform. 2020;doi:10.1108/ACI-09-2020-0059.

\bibitem{Satrio2021}
Aditya~Satrio CB, Darmawan W, Nadia BU, Hanafiah N.
\newblock Time series analysis and forecasting of coronavirus disease in
  {Indonesia} using {ARIMA} model and {PROPHET}.
\newblock Procedia Comput Sci. 2021;179:524--532.
\newblock doi:10.1016/j.procs.2021.01.036.

\bibitem{Johnson2018}
Johnson LR, Gramacy RB, Cohen J, Mordecai E, Murdock C, Rohr J, et~al.
\newblock Phenomenological forecasting of disease incidence using
  heteroskedastic {Gaussian} processes: {A} dengue case study.
\newblock Ann Appl Stat. 2018;12(1):27--66.
\newblock doi:10.1214/17-AOAS1090.

\bibitem{Velasquez2020}
Arias~Velásquez RM, Mejía~Lara JV.
\newblock Forecast and evaluation of {COVID}-19 spreading in {USA} with
  reduced-space {Gaussian} process regression.
\newblock Chaos Solitons Fractals. 2020;136:109924.
\newblock doi:10.1016/j.chaos.2020.109924.

\bibitem{Qian2020}
Qian Z, Alaa AM, van~der Schaar M.
\newblock In: Advances in {Neural} {Information} {Processing} {Systems}.
  vol.~33; 2020. p. 10729--10740.
\newblock Available from:
  \url{https://proceedings.neurips.cc/paper/2020/hash/79a3308b13cd31f096d8a4a34f96b66b-Abstract.html}.

\bibitem{Hawyrluk2021}
Hawryluk I, Hoeltgebaum H, Mishra S, Miscouridou X, Schnekenberg RP, Whittaker
  C, et~al.
\newblock Gaussian process nowcasting: {Application} to {COVID}-19 mortality
  reporting.
\newblock Proceedings of the 37th Conference on Uncertainty in Artificial
  Intelligence. 2021;161:1258--1268.
\newblock Available from:
  \url{https://proceedings.mlr.press/v161/hawryluk21a.html}.

\bibitem{Carpenter2017}
Carpenter B, Gelman A, Hoffman MD, Lee D, Goodrich B, Betancourt M, et~al.
\newblock Stan: {A} probabilistic programming language.
\newblock J Stat Softwb. 2017;76:1--32.
\newblock doi:10.18637/jss.v076.i01.

\bibitem{Gneiting2007}
Gneiting T, Balabdaoui F, Raftery AE.
\newblock Probabilistic forecasts, calibration and sharpness.
\newblock J R Stat Soc Series B. 2007;69(2):243--268.
\newblock doi:10.1111/j.1467-9868.2007.00587.x.

\bibitem{Bosse2023}
Bosse NI, Abbott S, Cori A, van Leeuwen E, Bracher J, Funk S. Transformation of
  forecasts for evaluating predictive performance in an epidemiological
  context.
\newblock medRxiv [Preprint]; 2023 [cited 2022 March 3].
\newblock Available from: \url{https://doi.org/10.1101/2023.01.23.23284722}.
  doi:10.1101/2023.01.23.23284722.

\bibitem{Bosse2022_Rpackage}
Bosse N, Abbott S, Gruson H, Bracher J, Funk S. Scoringutils: {Utilities} for
  {Scoring} and {Assessing} {Predictions}; 2022.
\newblock R package version 1.0.1.
\newblock Available from:
  \url{https://cran.r-project.org/web/packages/scoringutils}.

\bibitem{Funk2019}
Funk S, Camacho A, Kucharski AJ, Lowe R, Eggo RM, Edmunds WJ.
\newblock Assessing the performance of real-time epidemic forecasts: {A} case
  study of {Ebola} in the {Western} {Area} region of {Sierra} {Leone}, 2014-15.
\newblock PLoS Comput Biol. 2019;15(2):e1006785.
\newblock doi:10.1371/journal.pcbi.1006785.

\bibitem{Jordan2022}
Jordan A, Krueger F, Lerch S. {ScoringRules}: {Scoring} {Rules} for
  {Parametric} and {Simulated} {Distribution} {Forecasts}; 2022.
\newblock R package version 1.0.2.
\newblock Available from:
  \url{https://cran.r-project.org/web/packages/scoringRules}.

\bibitem{McDonald2021}
McDonald DJ, Bien J, Green A, Hu AJ, DeFries N, Hyun S, et~al.
\newblock Can auxiliary indicators improve {COVID}-19 forecasting and hotspot
  prediction?
\newblock Proc Natl Acad Sci U S A. 2021;118(51):e2111453118.
\newblock doi:10.1073/pnas.2111453118.

\bibitem{RCoreTeam2022}
{R Core Team}. R: {A} language and environment for statistical computing.
\newblock R Foundation for Statistical Computing; 2022.
\newblock Version 4.2.1.
\newblock Available from: \url{https://www.R-project.org/}.

\bibitem{Brauner2021}
Brauner JM, Mindermann S, Sharma M, Johnston D, Salvatier J, Gavenčiak T,
  et~al.
\newblock Inferring the effectiveness of government interventions against
  {COVID}-19.
\newblock Science. 2021;371(6531):eabd9338.
\newblock doi:10.1126/science.abd9338.

\bibitem{Banholzer2021}
Banholzer N, Weenen Ev, Lison A, Cenedese A, Seeliger A, Kratzwald B, et~al.
\newblock Estimating the effects of non-pharmaceutical interventions on the
  number of new infections with {COVID}-19 during the first epidemic wave.
\newblock PLoS One. 2021;16(6):e0252827.
\newblock doi:10.1371/journal.pone.0252827.

\bibitem{Cramer2022}
Cramer EY, Ray EL, Lopez VK, Bracher J, Brennen A, Castro~Rivadeneira AJ,
  et~al.
\newblock Evaluation of individual and ensemble probabilistic forecasts of
  {COVID}-19 mortality in the {United} {States}.
\newblock Proc Natl Acad Sci U S A. 2022;119(15):e2113561119.
\newblock doi:10.1073/pnas.2113561119.

\bibitem{Sherratt2022}
Sherratt K, Gruson H, Grah R, Johnson H, Niehus R, Prasse B, et~al.. Predictive
  performance of multi-model ensemble forecasts of {COVID}-19 across {European}
  nations.
\newblock medRxiv [Preprint]; 2022 [cited 2022 March 3].
\newblock Available from:
  \url{https://www.medrxiv.org/content/10.1101/2022.06.16.22276024v1}.
  doi:10.1101/2022.06.16.22276024.

\bibitem{Reinhart2021}
Reinhart A, Brooks L, Jahja M, Rumack A, Tang J, Agrawal S, et~al.
\newblock An open repository of real-time {COVID}-19 indicators.
\newblock Proc Natl Acad Sci U S A. 2021;118(51):e2111452118.
\newblock doi:10.1073/pnas.2111452118.

\bibitem{Schwabe2021}
Schwabe A, Persson J, Feuerriegel S.
\newblock Predicting COVID-19 spread from large-scale mobility data.
\newblock In: Proceedings of the 27th ACM SIGKDD Conference on Knowledge
  Discovery \& Data Mining; 2021. p. 3531–3539.
\newblock Available from: \url{https://doi.org/10.1145/3447548.3467157}.
  doi:10.1145/3447548.3467157.

\bibitem{Rodriguez2021}
Rodríguez A, Tabassum A, Cui J, Xie J, Ho J, Agarwal P, et~al.
\newblock {DeepCOVID}: {An} {Operational} {Deep} {Learning}-driven {Framework}
  for {Explainable} {Real}-time {COVID}-19 {Forecasting}.
\newblock In: Proceedings of the {AAAI} {Conference} on {Artificial}
  {Intelligence}. vol.~35; 2021. p. 15393--15400.
\newblock Available from:
  \url{https://ojs.aaai.org/index.php/AAAI/article/view/17808}.
  doi:10.1609/aaai.v35i17.17808.

\bibitem{Reiner2021}
Reiner RC, Barber RM, Collins JK, Zheng P, Adolph C, Albright J, et~al.
\newblock Modeling {COVID}-19 scenarios for the {United} {States}.
\newblock Nat Med. 2021;27(1):94--105.
\newblock doi:10.1038/s41591-020-1132-9.

\bibitem{Noh2021}
Noh J, Danuser G.
\newblock Estimation of the fraction of {COVID}-19 infected people in {U}.{S}.
  states and countries worldwide.
\newblock PLoS One. 2021;16(2):e0246772.
\newblock doi:10.1371/journal.pone.0246772.

\end{thebibliography}

\newpage

\section*{Acknowledgments}

We thank Anne Cori for her support with the implementation of EpiEstim. We thank Nikos Bosse for his suggestions on evaluating probabilistic forecasts.

\section*{Ethics approval}

Ethics approval was not required for this study. 

\section*{Competing interests}

SF reports grants from the Swiss National Science Foundation outside of the submitted work. All other authors declare no competing interests.

\section*{Funding}

NB and SF acknowledge funding from the Swiss National Science Foundation (SNSF) as part of the Eccellenza grant 186932 on "Data-driven health management".  TM, SM, HJTU and SB acknowledge funding from the MRC Centre for Global Infectious Disease Analysis (reference MR/R015600/1), jointly funded by the UK Medical Research Council (MRC) and the UK Foreign, Commonwealth and Development Office (FCDO), under the MRC/FCDO Concordat agreement and is also part of the EDCTP2 programme supported by the European Union; and acknowledges funding by Community Jameel. The funding bodies had no control over design, conduct, data, analysis, review, reporting, or interpretation of the research conducted. SB is funded by the National Institute for Health Research (NIHR) Health Protection Research Unit in Modelling and Health Economics, a partnership between the UK Health Security Agency, Imperial College London and LSHTM (grant code NIHR200908). Disclaimer: “The views expressed are those of the author(s) and not necessarily those of the NIHR, UK Health Security Agency or the Department of Health and Social Care.” SB acknowledges financial support from the Novo Nordisk Foundation via The Novo Nordisk Young Investigator Award (NNF20OC0059309) which also supports SM. SB acknowledges financial support from the Danish National Research Foundation via a chair grant. SB acknowledges financial support from The Eric and Wendy Schmidt Fund For Strategic Innovation via the Schmidt Polymath Award (G-22-63345).

\section*{Author Contributions}

NB, TM, SM, and SB conceptualized the study. NB collected and analyzed the data. NB, TM, SM, and SB contributed to methodology. All authors contributed to interpretation and visualization of data and results. NB wrote the original draft. All authors contributed to writing -- review \& editing.

\section*{Supporting information}

\textbf{S1 Appendix. Supplementary material.} Fig S1. Labeling of the epidemic trajectories. Fig S2. Forecasting performance for different modeling choices. Fig S3. Probabilistic forecasts for Arizona by model. Fig S4. Probabilistic forecasts for Illinois by model. Fig S5. Probabilistic forecasts for Maryland by model. Fig S6. Probabilistic forecasts for New Jersey by model. Fig S7. Probabilistic forecasts for New York by model. Fig S8. Evaluation of forecasting performance by US state. Fig S9. Evaluation of forecasting performance by epidemic phase. 

\clearpage
\setcounter{figure}{0}
\appendix
\doublespacing
\renewcommand{\thefigure}{S\arabic{figure}}
\title{\centering\LARGE\singlespacing{\textbf{S1 Appendix}}}

\sloppy
\raggedbottom

\begin{figure}[!htpb]
    \centering
    \includegraphics[width=\linewidth]{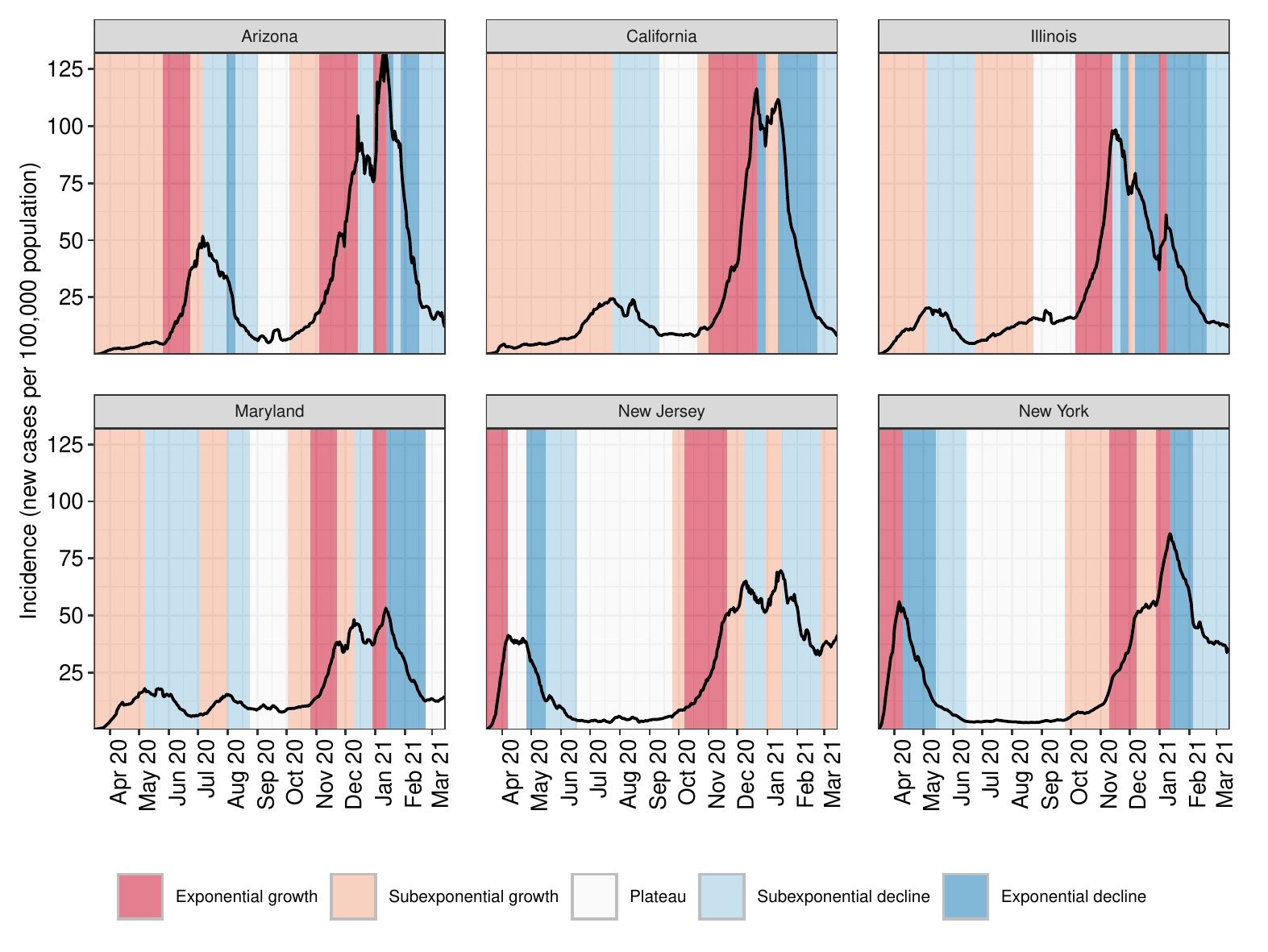}
    \caption{\textbf{Labeling of the epidemic trajectories.} Epidemic trajectories in US states manually labeled according to five phases: exponential increase, subexponential increase, plateau, subexponential decline, exponential decline.}
    \label{fig:phases}
\end{figure}

\begin{figure}[!htpb]
    \centering
    \includegraphics[width=16cm]{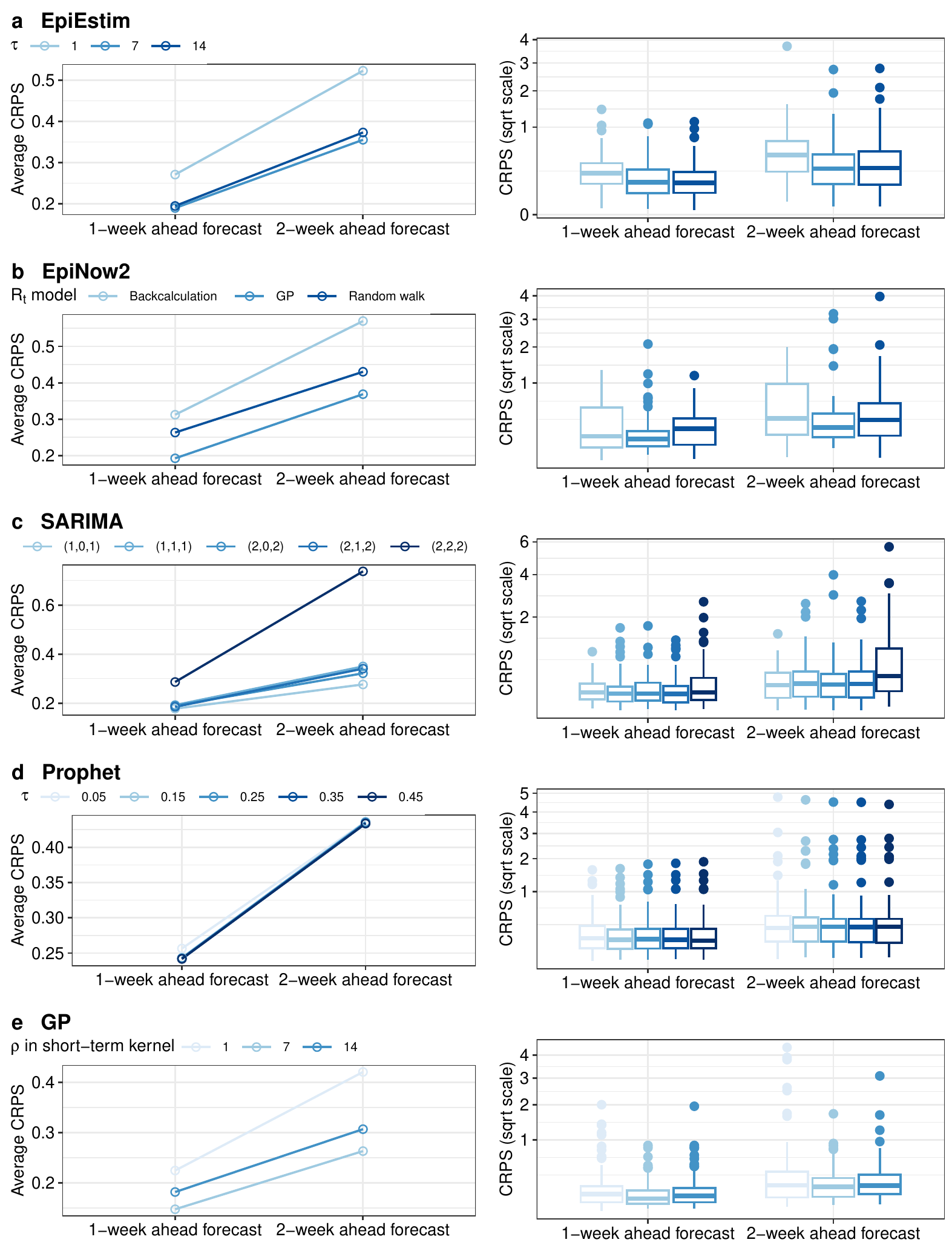}
    \caption{\textbf{Forecasting performance for different modeling choices.} CRPS (left average scores, right boxplot of scores) for the 1- and 2-week ahead forecast by model specification. \textbf{(a)}~EpiEstim. \textbf{(b)}~EpiNow2. \textbf{(c)}~SARIMA. \textbf{(d)}~Prophet. \textbf{(e)}~GP.}
    \label{fig:experiments}
\end{figure}

\begin{figure}
    \centering
    \includegraphics{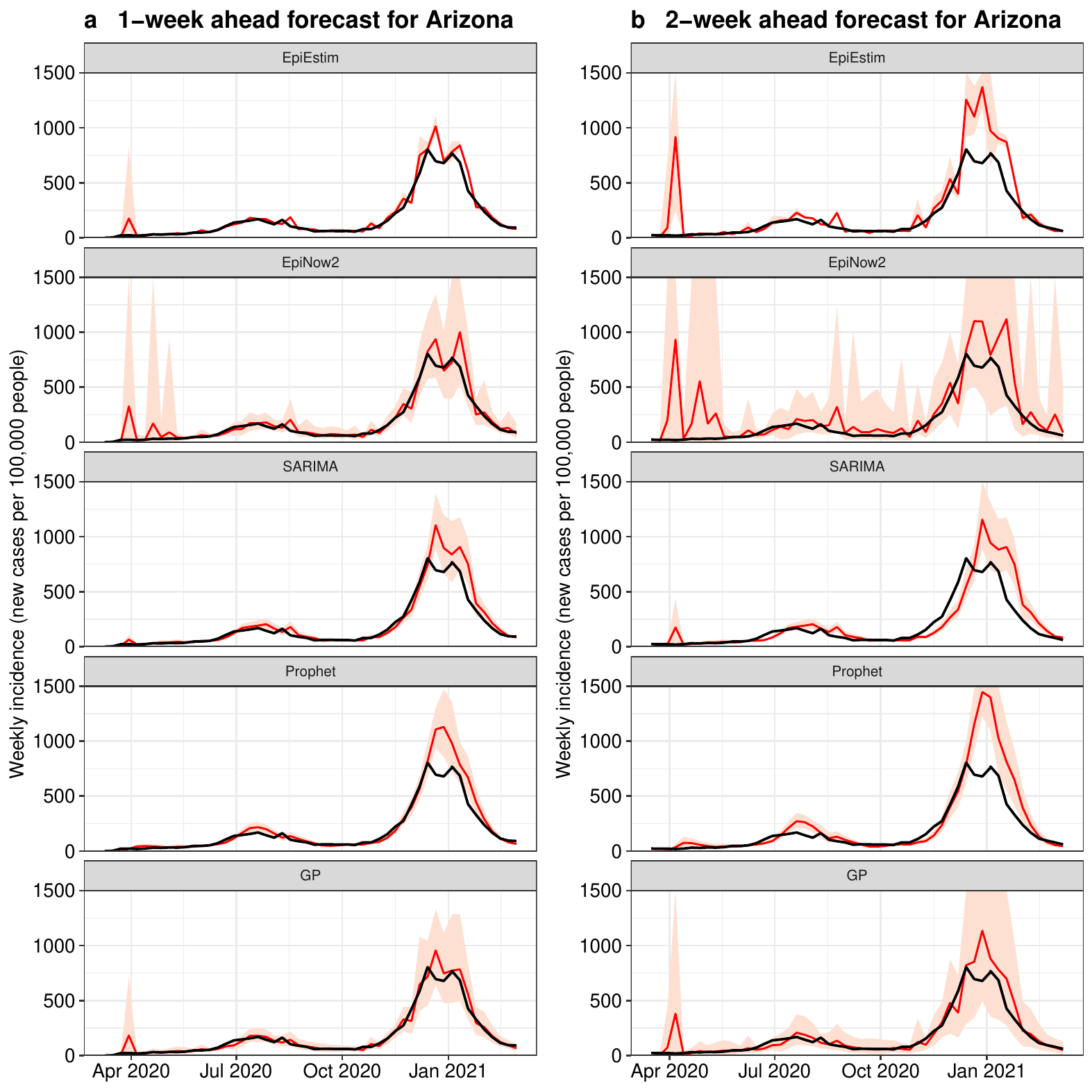}
    \caption{\textbf{Probabilistic forecasts for Arizona by model.} Probabilistic forecast (posterior mean as red line, 95\%-PI as shaded area) for Arizona by each model. \textbf{(a)}~1-week ahead forecast. \textbf{(b)}~2-week ahead forecast. Observed incidence is shown with black lines.}
    \label{fig:forecasts-az}
\end{figure}

\begin{figure}
    \centering
    \includegraphics{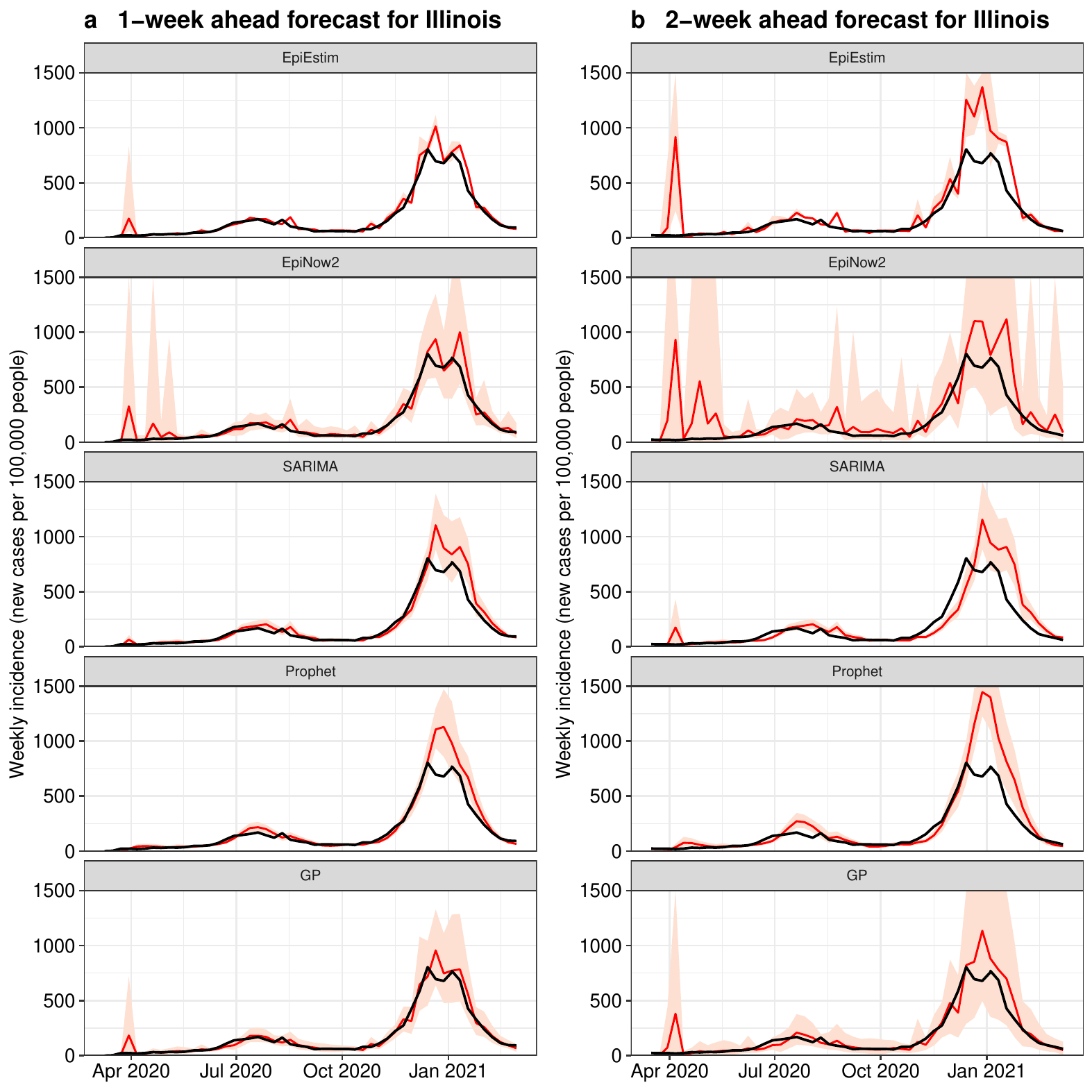}
    \caption{\textbf{Probabilistic forecasts for Illinois by model.} Probabilistic forecast (posterior mean as red line, 95\%-PI as shaded area) for Illinois by each model. \textbf{(a)}~1-week ahead forecast. \textbf{(b)}~2-week ahead forecast. Observed incidence is shown with black lines.}
    \label{fig:forecasts-il}
\end{figure}

\begin{figure}
    \centering
    \includegraphics{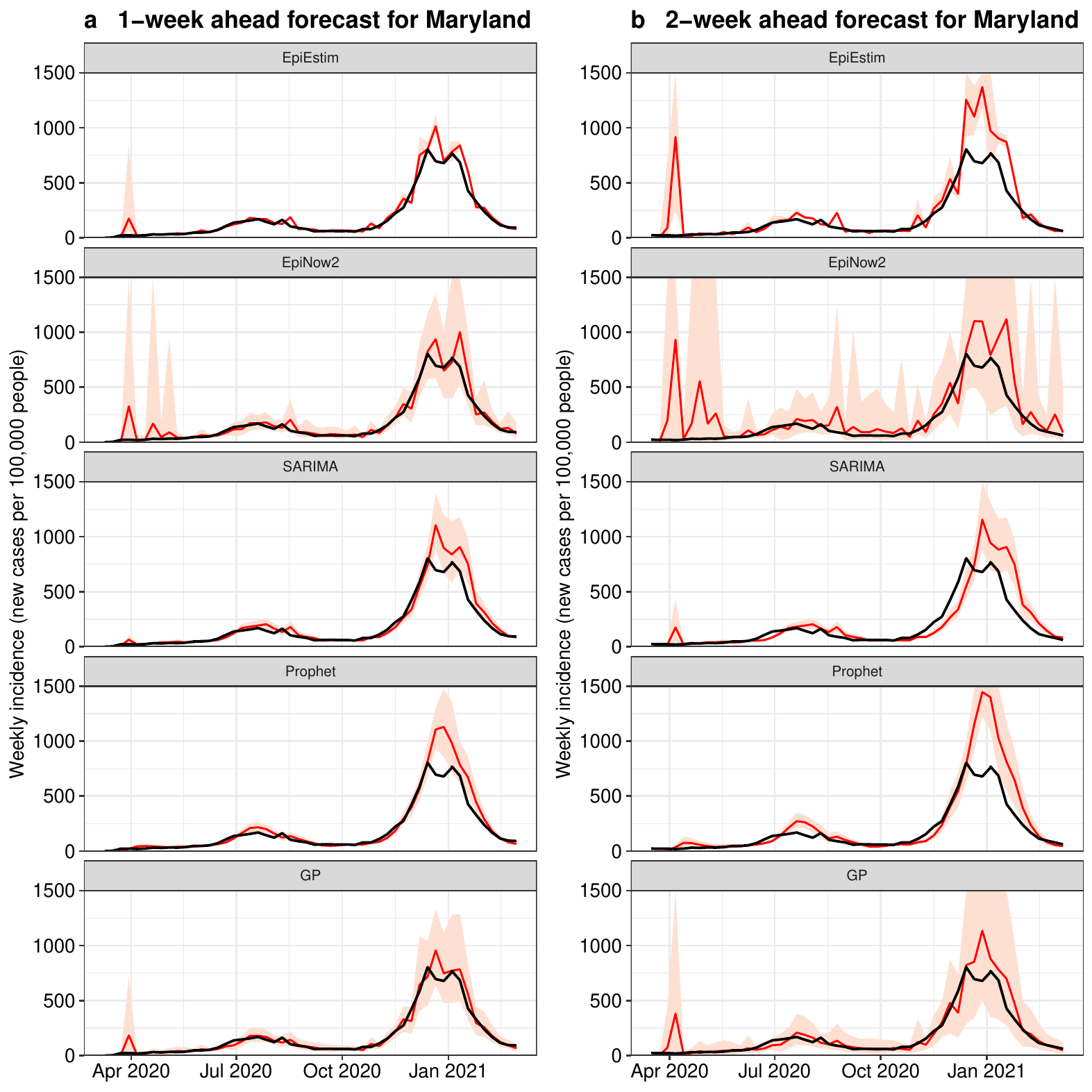}
    \caption{\textbf{Probabilistic forecasts for Maryland by model.} Probabilistic forecast (posterior mean as red line, 95\%-PI as shaded area) for Maryland by each model. \textbf{(a)}~1-week ahead forecast. \textbf{(b)}~2-week ahead forecast. Observed incidence is shown with black lines.}
    \label{fig:forecasts-md}
\end{figure}

\begin{figure}
    \centering
    \includegraphics{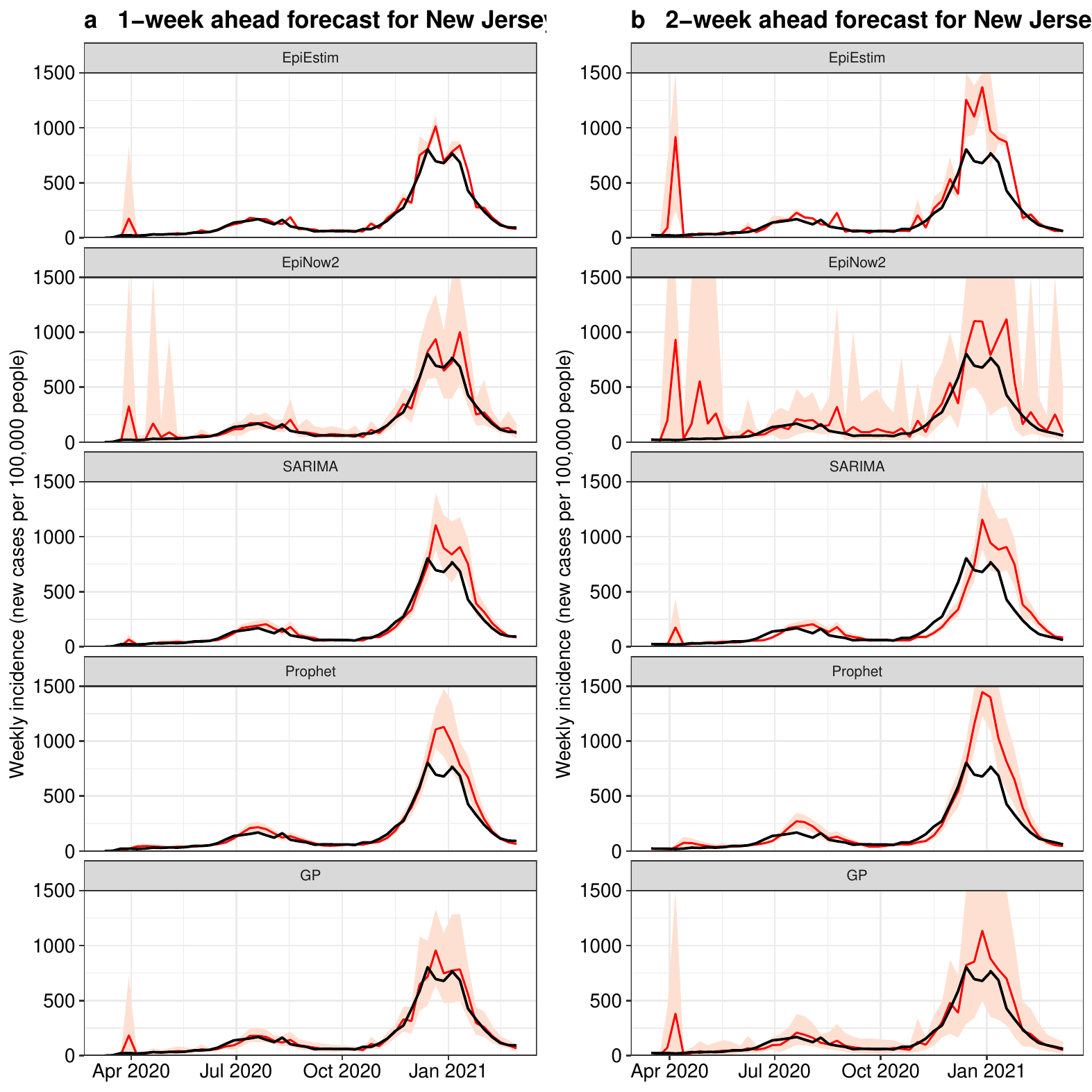}
    \caption{\textbf{Probabilistic forecasts for New Jersey by model.} Probabilistic forecast (posterior mean as red line, 95\%-PI as shaded area) for New Jersey by each model. \textbf{(a)}~1-week ahead forecast. \textbf{(b)}~2-week ahead forecast. Observed incidence is shown with black lines.}
    \label{fig:forecasts-nj}
\end{figure}
    
\begin{figure}
    \centering
    \includegraphics{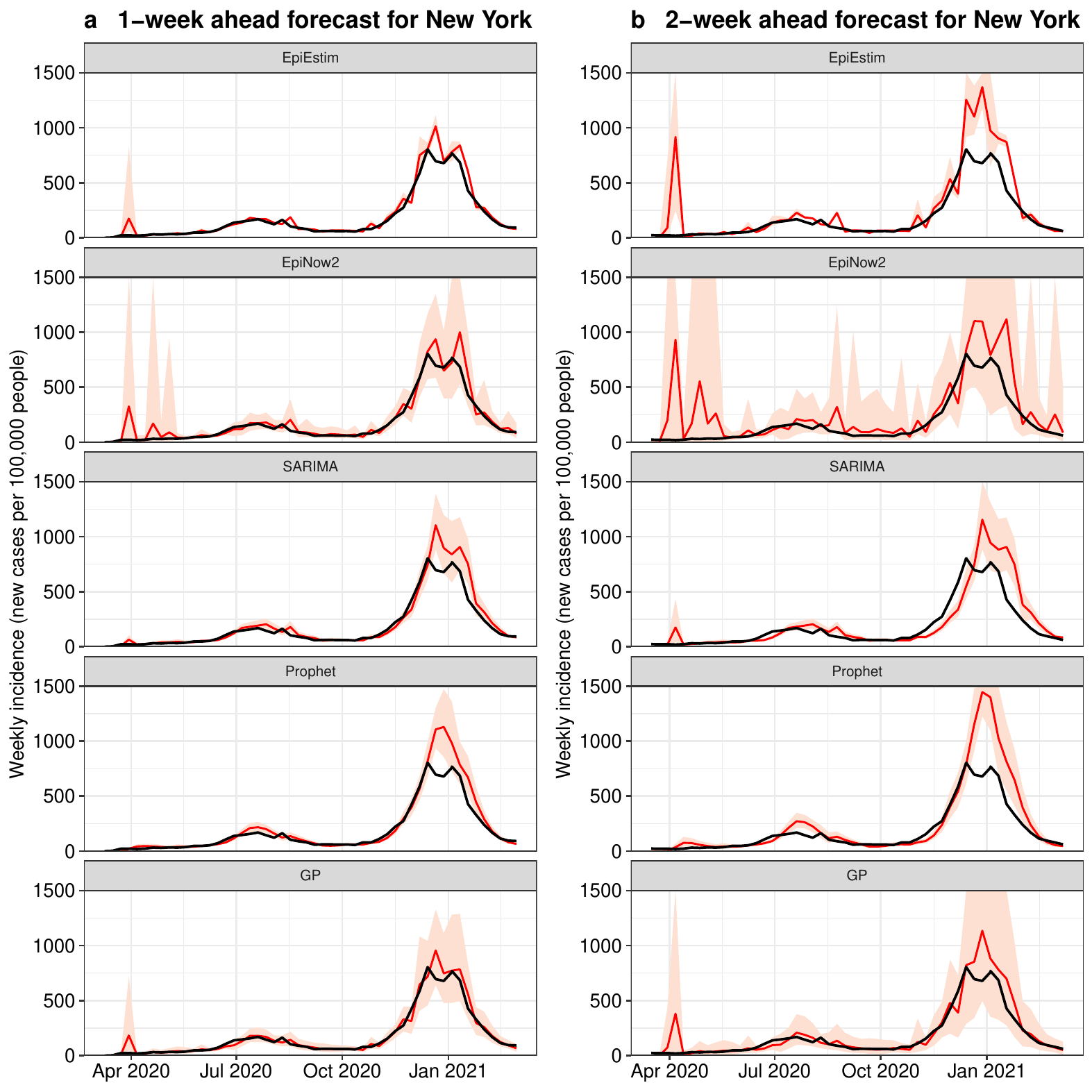}
    \caption{\textbf{Probabilistic forecasts for New York by model.} Probabilistic forecast (posterior mean as red line, 95\%-PI as shaded area) for New York by each model. \textbf{(a)}~1-week ahead forecast. \textbf{(b)}~2-week ahead forecast. Observed incidence is shown with black lines.}
    \label{fig:forecasts-ny}
\end{figure}

\begin{figure}[!htpb]
    \centering
    \includegraphics[width=\linewidth]{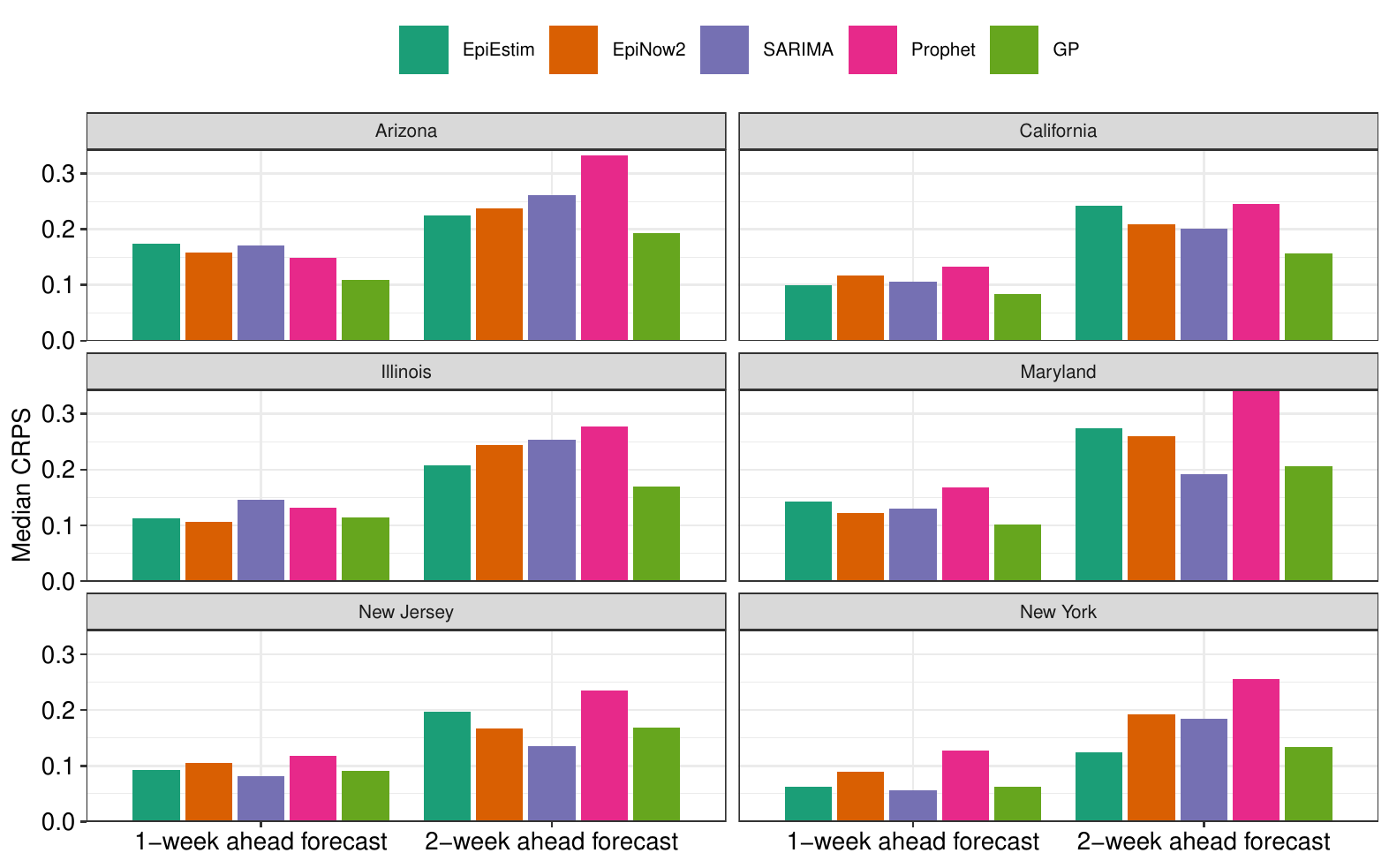}
    \caption{\textbf{Evaluation of forecasting performance by US state.} Median CRPS by US state for the 1- and 2-week ahead forecast by model.}
    \label{fig:crps_by_state}
\end{figure}

\begin{figure}[!htpb]
    \centering
    \includegraphics[width=\linewidth]{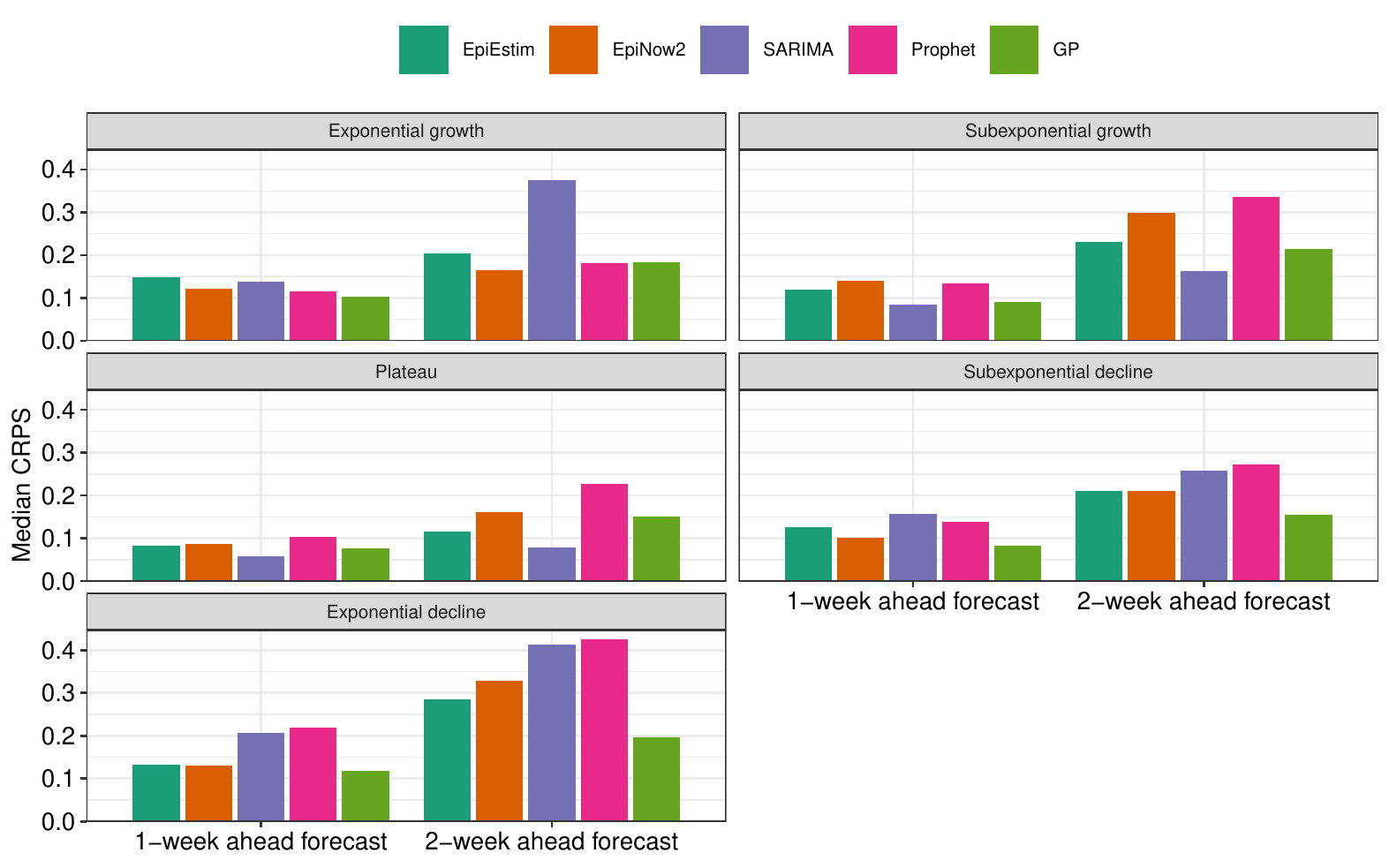}
    \caption{\textbf{Evaluation of forecasting performance by epidemic phase.} Median CRPS by different epidemic phases (see labeling in Figure~\ref{fig:phases}) for the 1- and 2-week ahead forecast by model.}
    \label{fig:crps_by_phase}
\end{figure}

\end{document}